\documentclass[10pt,twocolumn,twoside] {IEEEtran}
\usepackage{times}
\usepackage{subfigure}
\usepackage{latexsym,amsmath,amstext,amsfonts,amssymb,graphicx,epsfig}
\usepackage{algorithm, algorithmic, multirow, xcolor,booktabs}
\usepackage{epstopdf}
\usepackage{cite}
\usepackage{latexsym}
\usepackage{booktabs}
\usepackage{subfigure}
\usepackage{color}

\title{Knowledge-Aided STAP Using Low Rank and Geometry Properties}

\author{$^\ast$Zhaocheng~Yang,
Rodrigo~C.~de~Lamare,~\IEEEmembership{Senior Member~IEEE}, ~Xiang Li~\IEEEmembership{Member~IEEE} and Hongqiang Wang% <-this % stops a space
\thanks{Z. Yang, X. Li and H. Wang are with Research Institute of Space Electronics, Electronics Science and Engineering School, National University of Defense Technology, Changsha, 410073, China. e-mail: yangzhaocheng@gmail.com, lixiang01@vip.sina.com, oliverwhq@vip.tom.com.}% <-this % stops a space

\thanks{R. C. de Lamare is with Communications Research Group, Department of Electronics, University of York, YO10 5DD, UK. e-mail: rcdl500@ohm.york.ac.uk}% <-this % stops a space
}
\begin{document}
%\ninept
%
\maketitle

\begin{abstract}
This paper presents knowledge-aided space-time adaptive processing
(KA-STAP) algorithms that exploit the low-rank dominant clutter and
the array geometry properties (LRGP) for airborne radar
applications. The core idea is to exploit the fact that the clutter
subspace is only determined by the space-time steering vectors,
 {red}{where the Gram-Schmidt orthogonalization approach is employed
to compute the clutter subspace. Specifically, for a side-looking uniformly
spaced linear array, the}
algorithm firstly selects a group of linearly independent space-time
steering vectors using LRGP that can represent the clutter subspace.
By performing the Gram-Schmidt orthogonalization procedure, the
orthogonal bases of the clutter subspace are obtained, followed by
two approaches to compute the STAP filter weights. To overcome the
performance degradation caused by the non-ideal effects, a KA-STAP
algorithm that combines the covariance matrix taper (CMT) is
proposed. For practical applications, a reduced-dimension version of
the proposed KA-STAP algorithm is also developed. The simulation
results illustrate the effectiveness of our proposed algorithms, and
show that the proposed algorithms converge rapidly and provide a
SINR improvement over existing methods when using a very small
number of snapshots.
\end{abstract}

\section{Introduction}

Space-time adaptive processing (STAP) is considered to be an
efficient tool for detection of slow targets by airborne radar
systems in strong clutter environments
\cite{JWard1994,Klemm2006,Guerci2003,Melvin2004}. However, due to
the very high degrees of freedom (DoFs), the full-rank STAP has a
slow convergence and requires about twice the DoFs of the
independent and identically distributed (IID) training snapshots to
yield an average performance loss of roughly $3$dB \cite{JWard1994}.
In real scenarios, it is hard to obtain so many IID training
snapshots, especially in heterogeneous environments. Therefore, it
is desirable to develop STAP techniques that can provide high
performance in small training support situations.

Reduced-dimension and reduced-rank methods have been considered to
counteract the slow convergence of the full-rank STAP
\cite{JWard1994,Klemm2006,Guerci2003,Melvin2004,
Haimovich1997,Guerci2002,Wang1994,JScott1997,JScott1999,delamare_esp,RuiSTAP2010,fa2011}.
These methods can reduce the number of training snapshots to twice
the reduced dimension, or twice the clutter rank  {blue}{if we
assume that the degrees of freedom of the reduced dimension
correspond to the rank of the clutter}. The parametric adaptive
matched filter (PAMF) based on a multichannel autoregressive model
\cite{Roman2000} provides another alternative solution to the slow
convergence of the full-rank STAP. Furthermore, the sparsity of the
received data and filter weights has been exploited to improve the
convergence of a generalized sidelobe canceler architecture in
\cite{ZcYangTSP2011}. However, it still fundamental for radar
systems to improve the convergence performance of STAP algorithms or
reduce their required sample support in heterogeneous environments
because the number of required snapshots is large relative to those
needed in IID scenarios.

Recently developed knowledge-aided (KA) STAP algorithms have
received a growing interest and become a key concept for the next
generation of adaptive radar systems
\cite{R.Guerci2006,Rangaswamy2006,Gerlach2003,Bergin2006,Stoica2008,Xumin2011,
Ruifa2010, Melvin2006,
Xie2011,IvanW2010,Chen2008,Bidon2011,Tang2011,Wu2011}. The core idea
of KA-STAP is to incorporate prior knowledge, provided by digital
elevation maps, land cover databases, road maps, Global Positioning
System (GPS), previous scanning data and other known features, to
compute estimates of the clutter covariance matrix (CCM) with high
accuracy \cite{R.Guerci2006,Rangaswamy2006}. Among the previously
developed KA-STAP algorithms, there is a class of approaches that
exploit the prior knowledge of the clutter ridge to form the STAP
filter weights in \cite{Melvin2006,Xie2011, IvanW2010} and
\cite{Chen2008}. The authors in \cite{Melvin2006} introduced a
knowledge-aided parametric covariance estimation (KAPE) scheme by
blending both prior knowledge and data observations within a
parameterized model to capture instantaneous characteristics of the
cell under test (CUT). A modified sample matrix inversion (SMI)
procedure to estimate the CCM using a least-squares (LS) approach
has been described in \cite{Xie2011} to overcome the range-dependent
clutter non-stationarity in conformal array configurations. However,
both approaches require the pseudo-inverse calculation to estimate
the CCM and this often requires a computationally costly singular
value decomposition (SVD) \cite{Melvin2006}.
 {red}{Although two weighting approaches with lower
computations are discussed in \cite{Melvin2006}, they are suboptimal
approaches to the LSE by the SVD and the performance of these
approaches relative to the LSE by the SVD depends on
 {blue}{the radar system parameters, especially the array
characteristics \cite{Melvin2006}}.} Moreover, the latter approach
has not considered the situation when the prior knowledge has
uncertainties. Under the assumption of the known clutter ridge in
the angle-Doppler plane, the authors in \cite{IvanW2010} imposed the
sparse regularization to estimate the clutter covariance excluding
the clutter ridge. Although this kind of method can obtain a
high-resolution even using only one snapshot, it requires the
designer to know the exact positions of the clutter ridge resulting
in being sensitive to the prior knowledge. Furthermore, the
computational complexity caused by sparse recovery is expensive. A
data independent STAP method based on prolate spheroidal wave
functions (PSWF) has been considered in MIMO radar by incorporating
the clutter ridge \cite{Chen2008}, where the computational
complexity is significantly reduced compared with the approaches in
\cite{Melvin2006} and \cite{Xie2011}. However, it is highly
dependent on the ideal clutter subspace and is not robust against
clutter subspace mismatches.

In this paper, we propose KA-STAP algorithms using prior knowledge
of the clutter ridge that avoid the pseudo-inverse calculation,
require a low computational complexity, and mitigate the impact of
uncertainties of the prior knowledge.  {red}{Specifically, for a
side-looking uniformly spaced linear array (ULA),} the proposed
method selects a group of linearly independent space-time steering
vectors that can represent the ideal clutter subspace using prior
knowledge of the dominant low-rank clutter and the array geometry
properties (LRGP). The orthogonal bases of the clutter ideal
subspace are computed by a Gram-Schmidt orthogonalization procedure
on the selected space-time steering vectors. Two robust approaches
to compute the STAP filter weights are then presented based on the
estimated clutter subspace. To overcome the performance degradation
caused by the  {red}{internal} clutter motion (ICM), we employ a
covariance matrix taper (CMT) to the estimated CCM. The array
calibration methods discussed in \cite{Melvin2006} can be applied to
our proposed algorithm to mitigate the
 {blue}{impact} of non-ideal factors, such as channel
mismatching.
 Moreover, a reduced-dimension
version of the proposed KA-STAP algorithm is devised for practical
applications. Finally, simulation results demonstrate the
effectiveness of our proposed algorithms.

The main contributions of our paper are:

(i) A KA-STAP algorithm using  {blue}{prior knowledge} of the LRGP
is proposed for airborne radar applications.

(ii) A KA-STAP combining CMT is introduced to counteract the
performance degradation caused by ICM  {red}{and prior knowledge
uncertainty,}
 and a reduced-dimension
version is also presented for practical applications.
{red}{Furthermore, the proposed algorithm provides evidence for the
KAPE approach to directly use the received data and the calibrated
space-time steering vectors (only the spatial taper without the
temporal taper) to compute the assumed clutter amplitude.}

(iii) A detailed comparison is presented to show the computational
complexity of the proposed and existing algorithms.

(iv) A study and comparative analysis of our proposed algorithms
including the impact of inaccurate prior knowledge and non-ideal
effects on the SINR performance, the convergence speed and the
detection performance with other STAP algorithms is carried out.

The work is organized as follows. Section II introduces the signal
model in airborne radar applications. Section III details the
approach of the proposed KA-STAP algorithms and also discusses the
computational complexity. The simulated airborne radar data are used
to evaluate the performance of the proposed algorithms in Section
IV. Section V provides the summary and conclusions.

\section{Signal Model}

\begin{figure}[!htb]
\centering
\includegraphics[width=80mm]{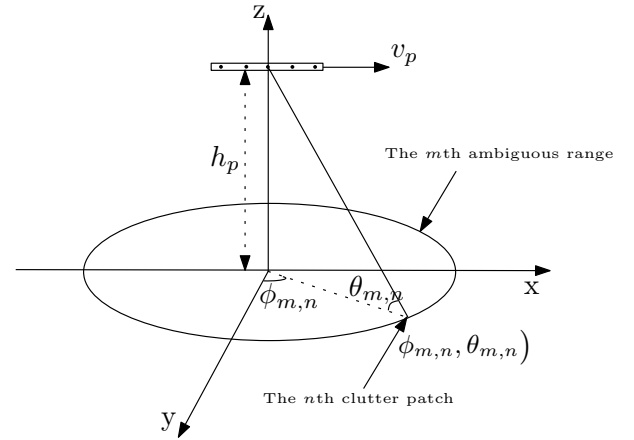}
\caption{
 Airborne radar geometry with a ULA antenna.} \label{radar_geometry_ula}
\end{figure}

The system under consideration is a side-looking pulsed Doppler
radar with a  {red}{ULA} consisting of $M$ elements on the airborne
radar platform, as shown in Fig.\ref{radar_geometry_ula}. The
platform is at altitude $h_p$ and moving with constant velocity
$v_p$. The radar transmits a coherent burst of pulses at a constant
pulse repetition frequency (PRF) $f_r=1/T_r$ , where $T_r$ is the
pulse repetition interval (PRI). The transmitter carrier frequency
is $f_c=c/\lambda_c$, where $c$ is the propagation velocity and
$\lambda_c$ is the wavelength. The number of pulses in a coherent
processing interval (CPI) is $N$. The received signal from the
iso-range of interest is represented by a space-time $NM\times1$
data vector ${\bf x}$.

 {red}{The received space-time clutter plus noise return
from a single range bin can be represented by} \cite{Melvin2004}
\begin{eqnarray}\label{model1}
    {\bf x} = \sum^{N_a}_{m=1}\sum^{N_c}_{n=1}\sigma_{m,n}
     {red}{{\bf v}(f_{s,m,n},f_{d,m,n}) \odot {\boldsymbol \alpha}(m,n)} + {\bf n},
\end{eqnarray}
where ${\bf n}$ is the Gaussian white thermal noise vector with the
noise power $\sigma^2_n$ on each channel and pulse; $N_a$ is the
number of range ambiguities; $N_c$ is the number of independent
clutter patches over the iso-range of interest; $f_{s,m,n}$ and
$f_{d,m,n}$ are the  {red}{spatial and Doppler frequencies} of the
$mn$th clutter patch, respectively;
 {red}{$\sigma_{m,n}$ is the complex amplitude for the
$mn$th clutter patch; ${\boldsymbol \alpha}(m,n)={\boldsymbol
\alpha}_d(m,n) \otimes {\boldsymbol \alpha}_s(m,n)$ is the
space-time random taper vector characterizing the voltage
fluctuation caused by  {blue}{non-ideal factors}, such as ICM and
channel mismatch (where ${\boldsymbol \alpha}_d(m,n)$ and
${\boldsymbol \alpha}_s(m,n)$ are the temporal and spatial random
tapers);} and ${\bf v}(f_{s,m,n},f_{d,m,n})$ is the $NM\times 1$
space-time steering vector for a clutter patch with
 {red}{$f_{s,m,n}$ and $f_{d,m,n}$}.

The space-time steering vector is given as the Kronecker product of
the temporal and spatial steering vectors, ${\bf
v}(f_{s,m,n},f_{d,m,n})={\bf v}_t(f_{d,m,n}) \otimes {\bf
v}_s(f_{s,m,n})$, which are given by \cite{JWard1994}
\begin{eqnarray}\label{model2}
    {\bf v}_t(f_{d,m,n})= [1,\cdots, \exp(j 2\pi (N-1)f_{d,m,n})]^T,
\end{eqnarray}
\begin{eqnarray}\label{model3}
    {\bf v}_s(f_{s,m,n})=[1, \cdots, \exp( j 2\pi(M-1)f_{s,m,n})]^T,
\end{eqnarray}
where $()^T$ denotes the transposition operation,
$f_{s,m,n}=\frac{d_a}{\lambda_c}\cos \theta_{m,n} \sin \phi_{m,n}$,
$f_{d,m,n}=\frac{2 v_p T_r}{\lambda_c} \cos \theta_{m,n}
\sin\phi_{m,n}$, and $d_a$ is the inter-sensor spacing of the ULA. If we stack all clutter patches' amplitudes into a vector
\begin{eqnarray}\label{model4}
    {\boldsymbol \sigma}=[\sigma_{1,1},\cdots,\sigma_{1,N_c},\cdots,\sigma_{N_a,1},\cdots,\sigma_{N_a,N_c}]^T,
\end{eqnarray}
 {red}{and assume there are no non-ideal factors,} then the
clutter plus noise received data denoted by (\ref{model1}) can be
{red}{described} as
\begin{eqnarray}\label{model5}
    {\bf x} = {\bf x}_{c} + {\bf n} = {\bf V}{\boldsymbol \sigma}  + {\bf n},
\end{eqnarray}
where ${\bf V}$ denotes the clutter space-time steering matrix,
given by
\begin{eqnarray}\label{model6}
\begin{split}
    {\bf V}= & [{\bf v}(\phi_{1,1},\theta_{1,1},f_{1,1}),\cdots, \\
    & {\bf v}(\phi_{1,N_c},\theta_{1,N_c},f_{1,N_c}), \cdots, \\
    & {\bf v}(\phi_{N_a,1},\theta_{N_a,1},f_{N_a,1}), \cdots,\\
    & {\bf v}(\phi_{N_a,N_c},\theta_{N_a,N_c},f_{N_a,N_c})].
\end{split}
\end{eqnarray}
Thus, the CCM based on (\ref{model5}) can be expressed as
\begin{eqnarray}\label{model7}
    {\bf R}_{c} = {\bf V}{\boldsymbol \Sigma}{\bf V}^H,
\end{eqnarray}
where ${\boldsymbol \Sigma}=E[{\boldsymbol \sigma}{\boldsymbol
\sigma}^H]$. Under the condition that the clutter patches are
independent from each other, ${\boldsymbol \Sigma}={\rm diag}({\bf
a})$, ${\bf a}=[a_{1,1},a_{1,2},\cdots,a_{N_a, N_c}]^T$ and $a_{m,n}
= E[|{\sigma}_{m,n}|^2]$ ($m = 1,\cdots,N_a$, $n=1,\cdots,N_c$) for
the statistics of the clutter patches. Here, $E[\cdot]$ denotes the
expectation operator, ${\rm diag}({\bf a})$ stands for a diagonal
matrix with the main diagonal taken from the elements of the vector
${\bf a}$ and $()^H$ represents the conjugate transpose of a matrix.

The optimal filter weight vector on maximizing the output SINR for
the Gaussian distribution clutter which is given by the full-rank
STAP processor can be written as \cite{Melvin2004}
\begin{equation}\label{model8}
    {\bf w}_{\rm opt}=\mu {\bf R}^{-1} {\bf s},
\end{equation}
where $\mu$ is a constant which does not affect the SINR
performance, ${\bf s}$ is the $NM \times1$ space-time steering
vector in the target direction, and ${\bf R}=E[{\bf x}{\bf
x}^H]={\bf R}_{c}+\sigma^2_n{\bf I}$ is the clutter plus noise
covariance matrix (${\bf I}$ is the identity matrix).

\section{KA-STAP Algorithms Using LRGP}

In this section, we firstly review the method that estimates the CCM
using a LS technique in \cite{Melvin2006,Xie2011} and point out the
existing problems of this method. Then, we detail the design and the
computational complexity of the proposed KA-STAP algorithms using
LRGP.

\subsection{CCM estimated by LS}

In practice, prior knowledge of certain characteristics of the radar
system and the aerospace platform, such as platform heading, speed
and altitude, array normal direction, and antenna phase steering,
etc., can be obtained from the Inertial Navigation Unit (INU) and
the GPS data \cite{Melvin2006,Bidon2011}. In other words, we can
obtain the values of the number of range ambiguities $N_a$, the
platform velocity $v_p$, and the elevation angle $\theta$. Thus, we
can develop KA-STAP algorithms based on  {red}{these} prior
knowledge, e.g., the methods described in \cite{Melvin2006,Xie2011,
IvanW2010} and \cite{Chen2008}.  {red}{In reality, the clutter
consists of returns over a continuous geographical region, which we
divide into a discrete set of clutter patches for analytical and
computational convenience. The rest of the discussion is on the
issues associated with choosing the number of clutter patches $N_c$.
A} possible approach is to assume a value of $N_c$ and discretize
the whole azimuth angle evenly into $N_c$ patches for each range bin
\cite{Melvin2006,Xie2011}.  {red}{In addition, it usually ignores
range ambiguities, i.e., $N_a=1$, where the justification can be
seen in \cite{Melvin2006}.} Then, the parameter ${\boldsymbol
\sigma}$  {red}{in (\ref{model5})} can be estimated using the
observation data by solving the LS problem as follows
\cite{Melvin2006,Xie2011},
\begin{eqnarray}\label{ls5}
    \hat{{\boldsymbol \sigma}} = {\rm arg}\min_{{\boldsymbol \sigma}}\|{\bf x}-{\bf V}{\boldsymbol \sigma}\|^2,
\end{eqnarray}
where  {red}{$\hat{{\boldsymbol
\sigma}}=[\hat{{\sigma}}_{1},\hat{{\sigma}}_{2}, \cdots,
\hat{{\sigma}}_{N_c}]^T$.} Herein, the solution for the above
problem based on an LS technique is given by
\begin{eqnarray}\label{ls6}
    \hat{{\boldsymbol \sigma}} = \big[{\bf V}^H{\bf V}\big]^{-1}{\bf V}^H{\bf x}.
\end{eqnarray}
Because ${\boldsymbol \sigma}$ depends only on the clutter
distribution, it does not vary significantly with the range under
homogeneous clutter environments. Furthermore, to avoid the effect
of the target signal at CUT, the near range bins of the CUT are used
to estimate ${\boldsymbol \sigma}$ \cite{Xie2011}, which is given by
\begin{eqnarray}\label{ls7}
    \hat{{\sigma}}^2_{m,n} = \frac{1}{L}\sum^{L}_{l=1}|\hat{{\sigma}}_{m,n;l}|^2,
\end{eqnarray}
where $2L$ is the total number of the secondary data. Then, the
estimated CCM by the LS method (we call it least-squares estimator
(LSE) in the following) is
\begin{eqnarray}\label{ls8}
    \hat{\bf R}_{c} = {\bf V}\hat{\boldsymbol \Sigma}{\bf V}^H.
\end{eqnarray}
 {red}{Then the clutter plus noise covariance matrix is
estimated as
\begin{eqnarray}\label{ls9}
    \hat{\bf R} = \hat{\bf R}_{c} + \hat{\sigma}^2_n{\bf I},
\end{eqnarray}
where $\hat{\sigma}^2_n$ is the estimated noise power level which
can be collected by the radar receiver when the radar transmitter
operates in a passive mode \cite{Klemm2006}.} Finally, the STAP
filter weights can be computed according to (\ref{model8}) using
{red}{$\hat{\bf R}$ instead of ${\bf R}$}.

However, there are several aspects that should be noted. First, the
above approach requires the designer to  {red}{choose the suitable
azimuth angle $\phi$ and the suitable number $N_c$} of the clutter
patches, which are difficult to obtain in practice. The
 {red}{selection of $N_c$ and $\phi$ will affect the
space-time steering vectors of the clutter patches, which affects
the estimation accuracy of the estimated CCM.} Specifically, if the
assumed number of clutter patches $N_c>NM$, then $\big[{\bf V}^H{\bf
V}\big]^{-1}$ does not exist.  {red}{Second}, the computational
complexity of the terms $\big[{\bf V}^H{\bf V}\big]^{-1}$ is very
high, i.e., $O((N_c)^3)+O(N_c(NM)^2)$, which should be avoided in
practice.  {red}{Two weighting approaches with lower computations
are discussed in \cite{Melvin2006}. {blue}{However, the solutions
obtained by the weighting approaches are suboptimal approximations
to the LSE obtained by the SVD. The performance of these approaches
relative to the LSE computed by the SVD depends on the radar system
parameters, especially the array characteristics}
\cite{Melvin2006}.}
 {blue}{In the presence of non-ideal factors in the clutter
component and despite the inclusion of the estimated
angle-independent channel mismatch in the space-time steering
vectors ${\bf V}$ and the use of the modified ${\bf V}$ to solve the
problem (\ref{ls5}), the techniques do not consider the impact of
the temporal random taper ${\boldsymbol \alpha}_d$. Nevertheless,
the received data vector ${\bf x}$ is formed by all non-ideal
factors. Thus, whether it is suitable to compute the parameter
${\boldsymbol \sigma}$ only considering the spatial random taper is
worth being investigated, as will be discussed in Section III.C.}

\subsection{Proposed KA-STAP Algorithm}

To overcome the rank-deficiency and the inverse of the matrix ${\bf
V}^H{\bf V}$, in the following, we will detail the proposed KA-STAP
algorithm to estimate the CCM using prior knowledge of LRGP.
 {red}{In this subsection, we only consider the ideal case of
the received data, i.e., the signal model in (\ref{model5}).}

From (\ref{model5}), we know that the clutter return is a linear combination of returns from all clutter patches. Thus, we have
\begin{eqnarray}\label{ebka1}
    {\rm span}({\bf R}_{c}) = {\rm span}({\bf V}) = {\rm span}({\bf V}{\bf V}^H).
\end{eqnarray}
\textit{Proof:} The first equation can be obtained from
(\ref{model7}). With regard to the second equation, let us denote
the SVD of the matrix ${\bf V}$ by  {red}{${\bf V}={\bf U}{\bf
C}{\bf D}^H$}. Then, we have
\begin{eqnarray}\label{ebka1.1}
\begin{split}
    & {\bf U}^H{\bf V}{\bf D}\big({\bf U}^H{\bf V}{\bf D}\big)^H = {\bf C}{\bf C}^H = \tilde{\bf C} \\
    \Rightarrow & {\bf U}^H\big({\bf V}{\bf V}^H\big){\bf U}=\tilde{\bf C},
\end{split}
\end{eqnarray}
where $\tilde{\bf C}={\bf C}{\bf C}^H$ is a real-valued diagonal
matrix. Thus, ${\bf U}$ is the orthogonal basis of the matrix ${\bf
V}{\bf V}^H$, i.e., ${\rm span}({\bf V}) = {\rm span}({\bf V}{\bf
V}^H)$.

Note that the orthogonal basis of the clutter subspace ${\bf U}$ can
be calculated by ${\bf V}$, or ${\bf V}{\bf V}^H$, herein we will
not need to compute that via the CCM. From (\ref{ebka1}), it also
results that the clutter subspace is independent from the power of
the clutter patches and is only determined by the clutter space-time
steering vectors. Moreover, from the above subsection, it is seen
that the clutter space-time steering vectors can be obtained using
the prior knowledge from the INU and GPS data. Therefore, it is
easier to compute the orthogonal bases of the clutter subspace ${\bf
U}$ by ${\bf V}$, or ${\bf V}{\bf V}^H$ than that by the CCM due to
the unknown power of the clutter patches. The other problem to
calculate the clutter subspace arising is that one should know the
clutter rank first. Fortunately, some rules for estimating the
clutter rank was discussed in previous literature, such as
\cite{JWard1994,Klemm2006,Zhang1997} and \cite{Nathan2007}.
Specially, for a side-looking ULA, the estimated clutter rank is
 {red}{approximated} by Brennan's rule as
\begin{eqnarray}\label{ebka2}
    {\rm rank}({\bf R}_{c}) \approx N_r=  {red}{\lceil M + \beta(N-1) \rceil},
\end{eqnarray}
where $\beta = 2v_pT_r/d_a$ and the brackets $\lceil \rceil$
indicate rounding to the  {red}{nearest largest integer}. In
\cite{Nathan2007}, this rule has been extended to the case with
arbitrary arrays. Usually, $N_r \ll NM$ and the STAP algorithms can
be performed in a low dimensional space so that the computational
complexity and the convergence can be significantly improved
\cite{Chen2008}. After the clutter rank is determined, there are
several approaches to compute the orthogonal bases of the clutter
subspace.

First, we can use the Lanczos algorithm \cite{Xu1994} applied to
${\bf V}{\bf V}^H$ to compute the clutter subspace eigenvectors. The
computational complexity of that using the Lanczos algorithm is on
the order of $O((NM)^2N_r + N^2_rNM) \ll O(N^3_c + N^2_cNM)+O(N_c(NM)^2)$. Moreover,
the computational complexity can be significantly reduced for the
case of ULA and constant PRF by exploiting the
Toepliz-block-Toeplitz structure of ${\bf V}{\bf V}^H$
\cite{Xu1994}.

Second, an alternative low-complexity approach is to perform the
Gram-Schmidt orthogonalization procedure on the space-time steering
vectors ${\bf V}$,  {red}{where the implementation steps of the
Gram-Schmidt orthogonalization are listed in Table \ref{tabel.ebka}
and interested readers are referred to \cite{Horn1985} for further
details. Note that this procedure is at the computational cost of
$O(\frac{(N_c+1)N_cNM}{2}+N^2_rNM) \ll O(N^3_c +
N^2_cNM)+O(N_c(NM)^2)$. It should be also noted that the approach of
the Gram-Schmidt orthogonalization can be applied to arbitrary
arrays if we can obtain the prior knowledge of the array geometry,
some radar system parameters and some information of the platform. }

 {red}{In particular, for the case of side-looking ULA, we
can further reduce the computational complexity to compute the
clutter subspace eigenvectors. Since the} dimension of the columns
of ${\bf V}$ should satisfy $N_c \gg N_r$, if we carry out the
Gram-Schmidt orthogonalization procedure on the columns of ${\bf V}$
one by one, this will result in unnecessary computations due to the
linear correlation among the columns. Thus, it is desirable to
directly find a group of vectors that are linear independent or
nearly linear independent (i.e., most of the vectors are linearly
independent and only very few vectors are linearly correlated).
 {red}{Fortunately,} for a ULA we have the following
proposition.

\textit{Proposition 1:} For the case of side-looking ULA and
constant PRF, the clutter subspace belongs to the subspace computed
by a group of space-time steering vectors $\{\bar{\bf
v}_p\}^{N_r}_{p=0}$, which are given by
\begin{eqnarray}\label{ebka3}
    \bar{\bf v}_p(n,m) = \exp( j 2\pi f_s (\beta n + m)),
\end{eqnarray}
where
\begin{eqnarray}\label{ebka4}
    f_s = \frac{p}{N_r}, p=0,1,\cdots,N_r-1.
\end{eqnarray}

\textit{Proof:} Let us stack the above space-time steering vectors into a $N_r \times NM$ matrix $\tilde{\bf V}$, which is shown as
\begin{eqnarray}\label{ebka5}
    \tilde{\bf V} = \left[\begin{array}{rccl} 1 & 1 & \cdots & 1 \\
    z_{0,0} & z_{0,1} & \cdots & z_{N,M} \\
    \vdots & \vdots & \vdots & \vdots \\
    z^{N_r-1}_{0,0} & z^{N_r-1}_{0,1} & \cdots & z^{N_r-1}_{N,M}
    \end{array}\right],
\end{eqnarray}
where
\begin{eqnarray}\label{ebka6}
    z_{n,m} = \exp( j 2\pi \frac{\beta n + m}{N_r}).
\end{eqnarray}
Note that $\tilde{\bf V}$ is a Vandermonde matrix of dimension $N_r
\times NM$.  {red}{For $z_{n,m}$, $n=0,\cdots,N-1$ and
$m=0,\cdots,M-1$, the number of linearly independent columns of
$\tilde{\bf V}$ is determined by the number of distinct values of
$\beta n + m$. If $\beta$ is an integer, the number of distinct
values of $\beta n + m$ is $N_r = \beta (N-1) + M$. If $\beta$ is a
rational value (not an integer), the number of distinct values of
$\beta n + m$ is larger than $N_r = \left\lceil M + \beta(N-1)
\right\rceil$.} Therefore, $\tilde{\bf V}$ has full rank, which is
equal to \cite{JWard1994}
\begin{eqnarray}\label{ebka7}
    {\rm rank}(\tilde{\bf V}) = \min (N_r, NM) = N_r.
\end{eqnarray}
The dimension of the clutter subspace is also $N_r$. Herein, the
clutter subspace shares the same subspace with $\tilde{\bf V}$. We
can then compute the clutter subspace by taking the Gram-Schmidt
orthogonalization procedure on the rows of $\tilde{\bf V}$.
Moreover, it should be noted that the computational complexity of
the second approach is on the order of
{red}{$O(\frac{(N_r+1)N_rNM}{2} + N^2_rNM)\ll
O(\frac{(N_c+1)N_cNM}{2}+N^2_rNM) \ll O(N^3_c +
N^2_cNM)+O(N_c(NM)^2$,} which exhibits a much lower complexity
compared with the LSE resulting in a very useful tool for practical
applications. It also avoids the procedure to determine the values
of the number of clutter patches $N_c$ and the azimuth angle $\phi$.

After computing the orthogonal basis of the clutter subspace, we try
to design the STAP filter weights by two different kinds of methods.
One is to use the minimum norm eigen-canceler (MNE) derived in
\cite{Haimovich1997} to form the filter weights. Specifically, the
MNE method is a linearly constrained beamformer with a minimum norm
weight vector appearing orthogonal to the clutter subspace, which is
described by \cite{Haimovich1997}
\begin{eqnarray}\label{ebka5}
\begin{split}
    &\quad\min_{\bf w}\quad{\bf w}^H{\bf w},\\
    {\rm subject \,\, to} \quad &{\bf U}^H{\bf w}=0 \quad {\rm and} \quad {\bf w}^H{\bf s}=1,
\end{split}
\end{eqnarray}
The solution to the above optimization problem in (\ref{ebka5}) is
provided by \cite{Haimovich1997}
\begin{eqnarray}\label{ebka6}
    \hat{\bf w}=\frac{\big({\bf I}-{\bf U}{\bf U}^H\big){\bf s}}{{\bf s}^H\big({\bf I}-{\bf U}{\bf U}^H\big){\bf s}}.
\end{eqnarray}
The other method tries to design the filter weights using both the
orthogonal bases of the computed clutter subspace and the
observation data. Let us first calculate the root-eigenvalues by
projecting the data on the clutter subspace ${\bf U}$, formulated as
\begin{eqnarray}\label{ebka7}
    \hat{\boldsymbol \gamma}={\bf U}^H{\bf x},
\end{eqnarray}
Then, the clutter plus noise covariance matrix $\hat{\bf R}$ can be estimated by
\begin{eqnarray}\label{ebka8}
    \hat{\bf R}={\bf U}\hat{\boldsymbol \Gamma}{\bf U}^H + \hat{\sigma}^2_n{\bf I},
\end{eqnarray}
where $\hat{\boldsymbol \Gamma}={\rm diag}(\hat{\boldsymbol \gamma}
\odot \hat{\boldsymbol \gamma}^\ast)$ and $\odot$ denotes the
Hadamard product. Finally, the STAP filter weights can be computed
by
\begin{eqnarray}\label{ebka9}
    \hat{\bf w} = \mu {\bf U} \big(\hat{\boldsymbol \Gamma}^{-1} + \frac{1}{\hat{\sigma}^2_n}{\bf I}\big){\bf U}^H{\bf s},
\end{eqnarray}
where we use the fact that $\hat{\bf R}^{-1}={\bf U}
\big(\hat{\boldsymbol \Gamma}^{-1} + \frac{1}{\hat{\sigma}^2_n}{\bf
I}\big){\bf U}^H$. The whole procedure of the proposed KA-STAP
algorithm is summarized in Table \ref{tabel.ebka}.

\begin{table}[!ht]
  \centering
  \caption{The Proposed KA-STAP Algorithm}\label{tabel.ebka}
  \small
  \begin{tabular}{|l|}
  \hline
  \textbf{Initialization:}\\
  $\beta = 2v_pT_r/d_a$, $N$, $M$,  {red}{$\hat{\sigma^2_n}$}.\\
  \hline
  \textbf{Select a group of space-time steering vectors}\\
  \quad $\{\bar{\bf v}_p\}^{N_r}_{p=0}$,\\
  \quad \textbf{where}\quad $\bar{\bf v}_p(n,m) = \exp( j 2\pi f_s (\beta n + m))$,\\
  \quad $N_r= M + \beta(N-1)$, \textbf{and}  $f_s = \frac{p}{N_r}, p=0,1,\cdots,N_r-1$,\\
  \hline
   {red}{
  \textbf{Compute calibrated space-time steering vectors}}\\
   {red}{\quad Estimate $\hat{\boldsymbol \alpha}_s$ using the methods in \cite{Melvin2006},}\\
   {red}{\quad where columns of $\hat{\boldsymbol \Xi}_s$ are all equivalent to
${\bf 1}_N \otimes \hat{\boldsymbol \alpha}_s$,}\\
   {red}{\quad ${\bf V}_s = {\bf V} \odot \hat{\boldsymbol \Xi}_s$,}\\
   {red}{\quad (In RD version, $\bar{\bf V}_s = {\bf S}^H_D{\bf V}_s$),}\\
  \hline
   {red}{\textbf{Compute ${\bf U}_s$}}\\
   {red}{\quad ${\bf u}_{s,0}=\bar{\bf v}_{s,0}/\|\bar{\bf v}_{s,0}\|$,}\\
   {red}{\quad $\tilde{\bf u}_{s,p}=\bar{\bf v}_{s,p} - \sum^{p-1}_{i=0} ({\bf u}^H_{s,i}\bar{\bf v}_{s,p}){\bf u}_{s,i}$,}\\
   {red}{\quad ${\bf u}_{s,p} = \tilde{\bf u}_{s,p}/\|\tilde{\bf u}_{s,p}\|, p=1,\cdots,N_r-1$,}\\
   {red}{\quad ${\bf U}_s = [{\bf u}_{s,0},\cdots,{\bf u}_{s,N_r-1}]$.}\\
   {red}{\quad (In RD version, $\bar{\bf U}_s$ instead of ${\bf U}_s$),}\\
  \hline
  \textbf{For each snapshot $l=1,\cdots,L$}\\
   {red}{\quad $\hat{\boldsymbol \gamma}_{l,s} ={\bf U}^H_s{\bf x}_l$,}\\
  \hline
   {red}{\textbf{Compute $\hat{\bf R}_c$}}\\
   {red}{\quad $\hat{\boldsymbol \Gamma}={\rm diag}(\frac{1}{L}\sum^L_{l=1}\hat{\boldsymbol \gamma}_{l,s} \odot \hat{\boldsymbol \gamma}^\ast_{l,s})$,}\\
   {red}{\quad $\hat{\bf R}_s = \hat{\bf U}_s\hat{\boldsymbol \Gamma}_s\hat{\bf U}^H_s$,}\\
   {red}{\quad Estimate $\hat{\bf T}_d$,}\\
   {red}{\quad $\hat{\bf R}_c = \hat{\bf R}_s \odot \hat{\bf T}_d$,}\\
   {red}{\quad (In RD version, $\bar{\bf U}_s$ instead of ${\bf U}_s$, $\hat{\bar{\bf T}}_d$ instead of $\hat{\bf T}_d$)}\\
  \hline
   {red}{\textbf{Compute ${\bf U}$}}\\
   {red}{\quad Adopt the Lanczos algorithm to $\hat{\bf R}_c$ to compute ${\bf U}$,}\\
   {red}{\quad (In RD version, $\hat{\bar{\bf R}}_c$ instead of $\hat{\bf R}_c$),}\\
  \hline
  \textbf{Filter weights computation}\\
  \quad $\hat{\bf w}=\frac{\big({\bf I}-{\bf U}{\bf U}^H\big){\bf s}}{{\bf s}^H\big({\bf I}-{\bf U}{\bf U}^H\big){\bf s}}$,\\
  \textbf{Or:}\\
  \quad $\hat{\bf w} = \mu {\bf U} \big(\hat{\boldsymbol \Gamma}^{-1} + \frac{1}{\hat{\sigma}^2_n}{\bf I}\big){\bf U}^H{\bf s}$.\\
   {red}{\quad (In RD version, $\bar{\bf U}$ instead of ${\bf U}$)}\\
  \hline
   \end{tabular}
\end{table}

\subsection{Proposed KA-STAP Employing CMT}

In practice, there are many non-ideal effects, such as the internal
clutter motion (ICM) and  {blue}{the channel mismatch}
\cite{Guerci2003}, which result in mismatch between the actual
clutter subspace and that computed by our proposed algorithm. In
this case, the performance of our proposed algorithm will
significantly degrade.  {red}{In the following, we will detail the
proposed KA-STAP employing CMT.}

 {red}{For the angle-dependent channel mismatch under normal circumstances,
 the transmit and receive antenna
patterns  {blue}{point} in the same direction and have a significant
maximum in the look direction.  {blue}{The energy from the sidelobes
is generally several orders of magnitude lower than that from the
mainbeam}. This will lead to clutter subspace leakage mainly coming
from the main beam \cite{Guerci2003}. Thus, the angle-dependent
channel mismatch can be approximated by spatial random tapers only
related to the main beam. Since the main beam is usually fixed in a
CPI, then this random tapers can be seen as angle-independent. For
the angle-independent channel mismatch, we assume the spatial taper
${\boldsymbol \alpha}_s$ is a random vector but stable over a CPI
due to the narrowband case considered in the paper. Herein, when in
presence of channel mismatch, the clutter plus noise
{blue}{received data vector} is given by \cite{Guerci2003}
\begin{eqnarray}\label{cmt_1}
    {\bf x} = ({\bf V} \odot {\boldsymbol \Xi}_s){\boldsymbol \sigma} + {\bf n},
\end{eqnarray}
where  {blue}{the} columns of ${\boldsymbol \Xi}_s$ are all
equivalent to ${\bf 1}_N \otimes {\boldsymbol \alpha}_s$ and ${\bf
1}_N$ denotes the all $1$ vector with dimension $N$. When
considering ICM, the received data can be represented as
\cite{Guerci2003}
\begin{eqnarray}\label{cmt_2}
    {\bf x} = ({\bf V}_s{\boldsymbol \sigma}) \odot
    ({\boldsymbol \alpha}_d \otimes {\bf 1}_M) + {\bf n},
\end{eqnarray}
where ${\bf V}_s = {\bf V} \odot {\boldsymbol \Xi}_s$ and
${\boldsymbol \alpha}_d$ is the temporal taper accounting for the ICM. Then,
the clutter plus noise covariance matrix is
\begin{eqnarray}\label{cmt_3}
    {\bf R} = {\bf R}_s \odot {\bf T}_d
     + \sigma^2_n{\bf I},
\end{eqnarray}
where
\begin{eqnarray}\label{cmt_4}
    {\bf R}_s = {\bf V}_s{\boldsymbol \Sigma}{\bf V}^H_s,
\end{eqnarray}
\begin{eqnarray}\label{cmt_5}
    {\bf T}_d = E[{\boldsymbol \alpha}_d{\boldsymbol \alpha}^H_d] \otimes {\bf 1}_{M,M},
\end{eqnarray}
where ${\bf T}_d$ denotes the space-time CMT accounting for the ICM and
${\bf 1}_{M,M}$ is the $M \times M$ all $1$ matrix. In order to obtain
the clutter plus noise covariance matrix, we should estimate ${\bf R}_s$
and ${\bf T}_d$ in (\ref{cmt_3}).}

 {red}{Regarding the estimation of ${\bf R}_s$,} we can
firstly use the array calibration methods discussed in
\cite{Melvin2006} to  {red}{estimate the spatial taper (denoted as
$\hat{\boldsymbol \alpha}_s$)}, which will not be discussed here due
to space limitations. The reader is
 {blue}{referred} to the literature \cite{Melvin2006} for
further details.  {red}{Then, substituting $\hat{\boldsymbol
\alpha}_s$ into ${\bf V}_s$, we obtain the
 {blue}{estimate} $\hat{\bf V}_s$. On the other hand, since
the elements of ${\boldsymbol \alpha}_d$ do not equate
 {blue}{to} zeros, we assume $\bar{\boldsymbol
\alpha}_d=[\frac{1}{\alpha_{d,1}},\cdots,\frac{1}{\alpha_{d,N}}]^T$.
If we multiply both side of (\ref{cmt_2}) by $\bar{\boldsymbol
\alpha}_d \otimes {\bf 1}_M$ and use the estimate $\hat{\bf V}_s$
instead of ${\bf V}_s$, then it becomes
\begin{eqnarray}\label{cmt_6}
    {\bf x}_s = {\bf x} \odot ({\boldsymbol \alpha}_d \otimes {\bf 1}_M) \approx
     \hat{\bf V}_s{\boldsymbol \sigma} + {\bf n}_s,
\end{eqnarray}
where ${\bf n}_s={\bf n} \odot ({\boldsymbol \alpha}_d \otimes {\bf
1}_M)$. In this situation, similarly as the analysis in Section
III.B, we can  {blue}{employ} the Gram-Schmidt orthogonalization
procedure to compute  {blue}{a matrix with} eigenvectors of
$\hat{\bf V}_s$, which is denoted as $\hat{\bf U}_s$. Then the
root-eigenvalues ${\boldsymbol \gamma}_s$ can be calculated by
\begin{eqnarray}\label{cmt_7}
    \hat{\boldsymbol \gamma}_s = \hat{\bf U}_s{\bf x}_s.
\end{eqnarray}
 {blue}{We can then} estimate ${\bf R}_s$ as
\begin{eqnarray}\label{cmt_8}
    \hat{\bf R}_s = \hat{\bf U}_s\hat{\boldsymbol \Gamma}_s\hat{\bf U}^H_s,
\end{eqnarray}
where
\begin{eqnarray}\label{cmt_9}
\begin{split}
    \hat{\boldsymbol \Gamma}_s &= {\rm diag}(\hat{\boldsymbol \gamma}_s \odot \hat{\boldsymbol \gamma}^\ast_s) \\
    & = {\rm diag}((\hat{\bf U}_s{\bf x}_s) \odot (\hat{\bf U}_s{\bf x}_s)^\ast) \\
    & = {\rm diag}((\hat{\bf U}_s \odot \hat{\bf U}^\ast_s) ({\bf x}_s \odot {\bf x}^\ast_s)) \\
    & = {\rm diag}((\hat{\bf U}_s \odot \hat{\bf U}^\ast_s) ({\bf x} \odot {\bf x}^\ast)) \\
    & = {\rm diag}((\hat{\bf U}_s {\bf x}) \odot (\hat{\bf U}_s{\bf x})^\ast)
\end{split}
\end{eqnarray}
Here, it uses the fact that the amplitude of the temporal taper
caused by the ICM is one. This fact can be seen the ICM models
 {blue}{reported in} \cite{JWard1994,Guerci2003}, which
will be also detailed afterwards. From (\ref{cmt_9}), we observe
that $\hat{\bf R}_s$ can be estimated using the received data ${\bf
x}$ directly without ${\boldsymbol \alpha}_d$. It also provides
evidence for the KAPE approach to directly use the received data and
the calibrated space-time steering vectors (only the spatial taper
without the temporal taper) to compute the parameter ${\boldsymbol
\sigma}$.}

 {red}{Regarding the estimation of ${\bf T}_d$, it can be obtained}
 via a rough knowledge of the interference environment (e.g., forest versus desert, bandwidth,
etc.)\cite{Guerci2002}. One common model, referred to as the
Billingsley model, is suitable for a land scenario. The only
parameters required to specify the clutter Doppler power spectrum
are essentially the operating wavelength and wind speed. The
operating wavelength is usually known, while the wind speed should
be estimated. Another common model, presented by J. Ward in
\cite{JWard1994}, is suitable for a water scenario. The temporal
autocorrelation of the fluctuations is Gaussian in shape with the
form:
\begin{eqnarray}\label{icm}
    \zeta(m)=\exp \big \{ - \frac{8\pi^2\sigma^2_vT^2_r}{\lambda^2_c}m^2\big\},
\end{eqnarray}
where $\sigma_v$ is the variance of the clutter spectral spread in
$m^2/s^2$. In the following simulations, we consider the CMT model
of the latter one.

 {red}{After computing the estimates $\hat{\bf R}_s$
and $\hat{\bf T}_d$, we can compute the CCM as $\hat{\bf R}_c = \hat{\bf R}_s \odot \hat{\bf T}_d$.
Since $\hat{\bf R}_c$ is still of low rank, we adopt
the Lanczos algorithm to compute the clutter subspace ${\bf U}$, where
the computational complexity is on the order of $O((NM)^2N'_r)$
($N'_r$ is the clutter rank of $\hat{\bf R}_c$).
Finally, the STAP filter weights are computed according to
(\ref{ebka6}) or (\ref{ebka9}). The whole procedure
can be seen in Table \ref{tabel.ebka}.}

\begin{figure*}[ht]
  \centering
  \subfigure[]{\label{prior.sub1}
  \includegraphics[width=7.8cm]{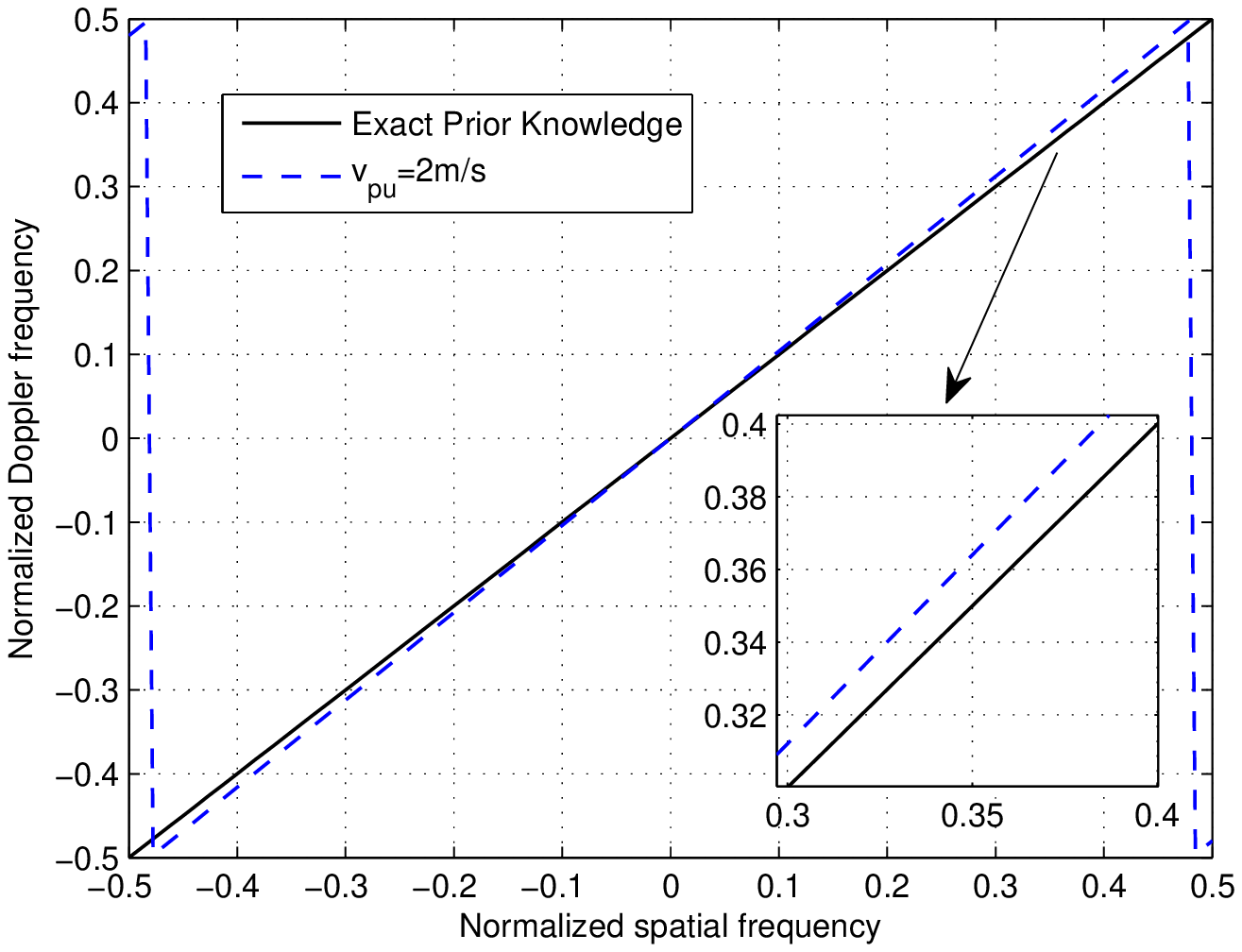}}
  \hspace{0.5in}
  \subfigure[]{\label{prior.sub2}
  \includegraphics[width=7.8cm]{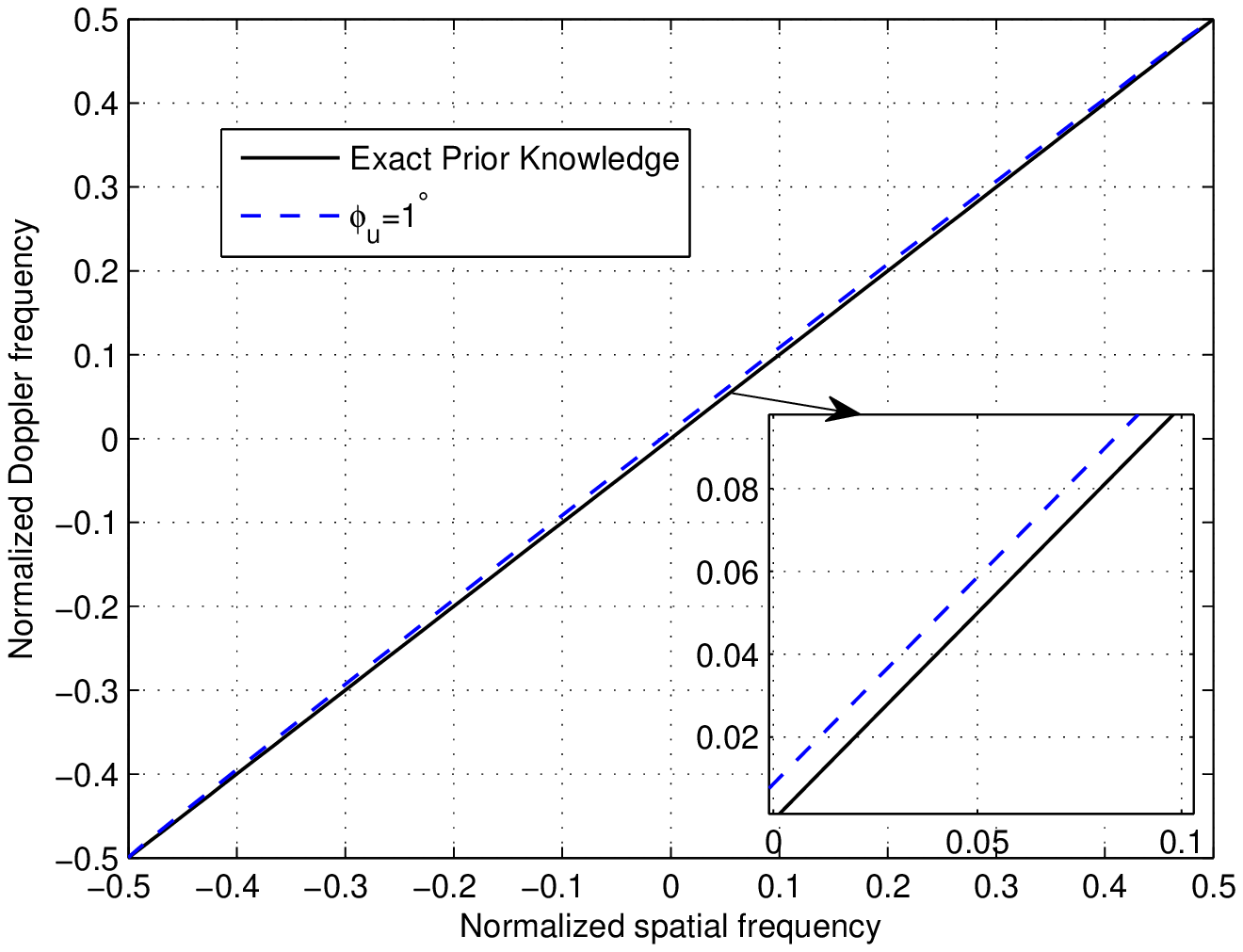}}
  %\hspace{0.5in}
  \caption{Impact of prior knowledge uncertainty to the clutter ridge in the spatio-temporal plane with (a) velocity deviation $v_{pu}=2$m/s and (b) $\phi_{u} = 1^\circ$.}\label{impactsprior}
\end{figure*}

 {red}{\textit{Prior knowledge uncertainty impact.} In the proposed algorithms,
the prior knowledge uncertainty, such as velocity misalignment and yaw angle misalignment, will have a great
impact on the performance. However, the scheme that employs the CMT will mitigate this impact.
To illustrate this, we take a typical airborne radar system for example. The parameters
of the radar system are listed at the beginning of Section IV. Consider a far field
scenario, the elevation angle will be close to zero resulting in $\cos \theta \approx 1$.
Let $v_{pu}$ and $\phi_u$ denote the velocity deviation and the yaw angle deviation, respectively.
Then, for a discretized azimuth angle $\phi$, the spatial frequency $f_s$ and Doppler frequency $f_d$
can be represented as
\begin{eqnarray}\label{f_s}
   f_s=\frac{d_a}{\lambda_c}\sin \phi,
\end{eqnarray}
\begin{eqnarray}\label{f_d}
    f_d = \frac{2(v_p+v_{pu})T_r}{\lambda_c}\sin(\phi + \phi_u).
\end{eqnarray}
From (\ref{f_s}) and (\ref{f_d}), we see that the prior knowledge
uncertainty will affect the position and shape of the clutter ridge,
which leads to the mismatch between the exact and the assumed
space-time steering vectors. Fig.\ref{impactsprior} provides a more
direct way to illustrate the impact of prior knowledge uncertainty
to the clutter ridge in the spatio-temporal plane. By employing a
CMT, the clutter spectra will become wider along the clutter ridge
in the figure including the exact clutter ridge. From this point of
view, the impact of prior knowledge uncertainty is mitigated.
Because the methods in \cite{Xie2011,IvanW2010} and \cite{Chen2008}
do not consider any strategies to mitigate the impact of prior
knowledge uncertainty, the performance will depend highly on the
accuracy of the prior knowledge. The KAPE approach in
\cite{Melvin2006} also adopts the CMT and can mitigate the impact of
prior knowledge uncertainty in a sense. But the differences between
the proposed algorithm and the KAPE approach lie in three aspects.
First, the KAPE approach estimates the CCM using the LS or some
approximate approaches. While the proposed algorithm estimates the
CCM using the Gram-Schmidt orthogonalization procedure (that is not
an approximate approach) by exploiting that the clutter subspace is
only determined by the space-time steering vectors. Furthermore, for
a side-looking ULA radar, the proposed algorithm directly selects a
group of linearly independent space-time steering vectors using the
LRGP and then takes the Gram-Schmidt orthogonalization procedure to
compute the clutter subspace. Second, the proposed algorithm shows
evidence  {blue}{that is feasible to directly use the received data
vector} and the calibrated space-time steering vectors (only the
spatial taper without the temporal taper) to compute the parameter
${\boldsymbol \sigma}$. Third, the proposed algorithm with an RD
version in the following section is presented to further reduce the
complexity. }

\subsection{Proposed Reduced-Dimension (RD) KA-STAP Algorithms}

From the above discussions, one aspect to be noted is that it is
impractical to use all the DoFs available at the ULA for reasons of
computational complexity when $NM$ is too large. In such situations,
a common approach is to break the full DoFs problem into a number of
smaller problems via the application of an $MN \times D$ (with
$D\ll MN$) transformation matrix ${\bf S}_D$ to the data
\cite{JWard1994}. Our proposed KA-STAP algorithms can be easily
extended to this kind of approach. By applying the reduced-dimension
transformation matrix ${\bf S}_D$ to the data and the space-time
steering vectors, we obtain
\begin{eqnarray}\label{rdka1}
    \bar{\bf x}={\bf S}^H_D{\bf x}, \quad \quad \bar{\bf V}={\bf S}^H_D{\bf V},
\end{eqnarray}
where $\bar{}$ denotes the results after the transformation. Then,
the reduced-dimension CCM $\bar{\bf R}_{c}$ becomes
\begin{eqnarray}\label{rdka2}
    \bar{\bf R}_{c}={\bf S}^H_D{\bf R}_{c}{\bf S}_D= \bar{\bf V} {\boldsymbol \Sigma}\bar{\bf V} = \bar{\bf U} \bar{\boldsymbol \Gamma} \bar{\bf U}^H.
\end{eqnarray}
In a manner similar to that of the proposed full-DoF KA-STAP
algorithm described in Section III.B, we compute the
orthogonal bases of the clutter subspace ${\bar{\bf U}}$, estimate
the CCM $\hat{\bar{\bf R}}_{c}={\bar{\bf U}} \hat{\bar{\boldsymbol
\Gamma}} {\bar{\bf U}}^H$, and then calculate the STAP filter
weights according to (\ref{ebka6}) or (\ref{ebka9}). When employing
a CMT to the ideal clutter covariance matrix, the final RD clutter
covariance matrix can be estimated as
\begin{eqnarray}\label{rdka3}
    \hat{\bar{\bf R}}_{c}= {red}{\hat{\bar{\bf R}}_s \odot \hat{\bar{\bf T}}_d
    = (\hat{\bar{\bf U}}_s \hat{\bar{\boldsymbol \Gamma}}_s \hat{\bar{\bf U}}^H_s) \odot \hat{\bar{\bf T}}_d,}
\end{eqnarray}
 {red}{where $\hat{\bar{\bf U}}_s$ is computed by taking
the Gram-Schmidt orthogonalization procedure to $\hat{\bar{\bf V}}_s={\bf S}^H_D\hat{\bf V}_s$,
$\hat{\bar{\boldsymbol \Gamma}}_s$ is calculated via (\ref{cmt_9}) using
$\bar{\bf x}$ and $\hat{\bar{\bf U}}_s$ instead of ${\bf x}$ and $\hat{{\bf U}}_s$,
and $\hat{\bar{\bf T}}_d$ denotes the estimated RD CMT. Again, the STAP
filter weights can be computed according to (\ref{ebka6}) or (\ref{ebka9}).}
By inspecting (\ref{rdka2}) and (\ref{rdka3}), we find that the computational
complexity of our proposed RD-KA-STAP algorithm is related to $D$
instead of $NM$ ($D \ll NM$), which leads to great computation
savings.

In this paper, we focus on the reduced-dimension technique known as
extended factored (EFA) algorithm or multibin element-space
post-Doppler STAP algorithm\cite{JWard1994}. The simulations with
this technique will show the performance of our proposed RD-KA-STAP
algorithm.

\subsection{Complexity Analysis}

Here we illustrate the computational complexity of the proposed
algorithms (shortened as LRGP KA-STAP and LRGP RD-KA-STAP) and other
existing algorithms, namely, the sample matrix inversion algorithm
(SMI), the EFA algorithm in \cite{JWard1994}, the
joint-domain-localized (JDL) algorithm in \cite{Wang1994}, the
CSMIECC algorithm in \cite{Xie2011}, and the KAPE algorithm in
\cite{Melvin2006}. In Table \ref{table.complexity}, $D$ denotes the
size of the reduced dimension. We can see that the computational
complexity of our proposed algorithms is significantly lower than
the CSMIECC and the KAPE algorithms  {red}{($N_r \ll N_c, NM$)},
which require the pseudo-inverse of the matrix ${\bf V}^H{\bf V}$.
With regard to the SMI algorithm, our proposed algorithms also show
a lower computational complexity because the number of snapshots
used for training the filter weights of the SMI is in the order of
$2NM$.

\begin{table*}[ht]
  \centering
  \caption{Computational Complexity of Algorithms}\label{table.complexity}
  \small
  %\footnotesize
  \begin{tabular}{|l|c|c|}
  \hline
  Algorithm & Estimate the CCM & Compute filter weights\\
  \hline
   SMI & $O\left(L(NM)^2\right)$ & $O\left((NM)^3 \right)$\\
  \hline
   EFA & $O\left(L(D)^2\right) + O\left(L\frac{N}{2}\log_2(N)\right)$ & $O\left(D^3 \right)$\\
  \hline
   JDL & $O\left(L(D)^2\right) + O\left(L\frac{NM}{2}\log_2(NM) \right)$ & $O\left(D^3 \right)$ \\
  \hline
   CSMIECC &  {red}{$O\left(L(NM)^2\right) + O\left(N_c(NM)^2\right)+ O\left(N^3_c+ N^2_cNM \right)$} & $O\left((NM)^3 \right)$ \\
  \hline
   KAPE &  {red}{$O\left(N_c(NM)^2\right)+ O\left(N^3_c + N^2_cNM\right)$} & $O\left((NM)^3 \right)$\\
  \hline
   LRGP KA-STAP &  {red}{$O\left(\frac{(N_r+1)N_rNM}{2} + N^2_rNM\right) + O\left(N_r(NM)^2\right)$} &  {red}{$O\left(N_r(NM)^2 \right)$}\\
  \hline
   LRGP RD-KA-STAP &  {red}{$O\left(\frac{(N_r+1)N_rD}{2} + N^2_rD\right) + O\left(N_rD^2\right)$} & $O\left(D^3 \right)$\\
  \hline
   \end{tabular}
\end{table*}

Although the computational complexity of the EFA and JDL algorithms
is lower than our proposed LRGP KA-STAP algorithm, two aspects
should be noted. One is that the number of snapshots used for
training filter weights is much larger than our proposed algorithms.
The other is that the computational complexity of EFA and JDL is
proportional to the number of Doppler frequencies of interest (we
only list the computation complexity for one Doppler frequency).
While our proposed algorithms only have to compute the CCM once for
different Doppler frequencies of interest. Besides, the
computational complexity of our proposed LRGP RD-KA-STAP is lower
than the EFA since $L$ in EFA is in the order of $2D$, where $D$ is
usually larger than $N_r$.

\section{Performance Assessment}

In this section, we assess the proposed KA-STAP algorithms by
computing the output SINR performance and probability of detection
performance using simulated radar data. The output SINR is defined
by
\begin{eqnarray}\label{eqsinr}
    {\rm SINR} = \frac{\left|\hat{\bf w}^H{\bf s}\right|^2}{\left|\hat{\bf w}^H{\bf R}\hat{\bf w}\right|}.
\end{eqnarray}
 {red}{Throughout the simulations, unless otherwise stated}, the simulated scenarios
use the following parameters: side-looking ULA, uniform
transmit pattern, $M=8$, $N=8$, $f_c=450$MHz, $f_r = 300$Hz, $v_p=50$m/s,
 {red}{$d_a=\lambda_c/2$,
$\beta = 1$, $N_r = \lceil M + \beta(N-1) \rceil = 15$,}
$h_p=9000$m, signal-to-noise ratio (SNR) of $0$dB, the
target located at $0^\circ$ azimuth with Doppler frequency $100$Hz,
clutter-to-noise ratio (CNR) of $50$dB, and unitary thermal noise
power. All presented results are averaged over $100$ independent
Monte Carlo runs.
\begin{figure*}[ht]
  \centering
  \subfigure[]{\label{icm.sub1}
  \includegraphics[width=7.8cm]{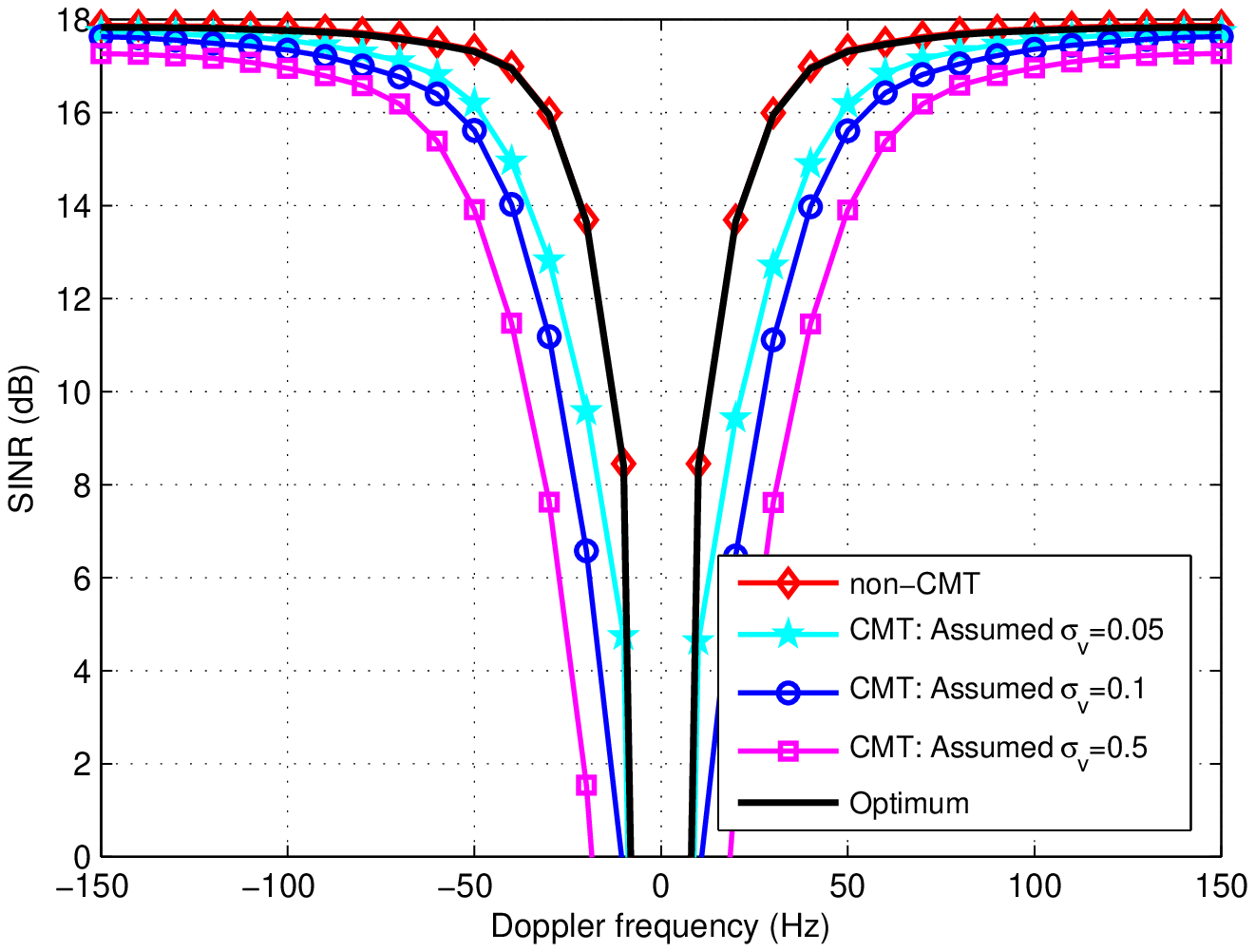}}
  \hspace{0.5in}
  \subfigure[]{\label{icm.sub2}
  \includegraphics[width=7.8cm]{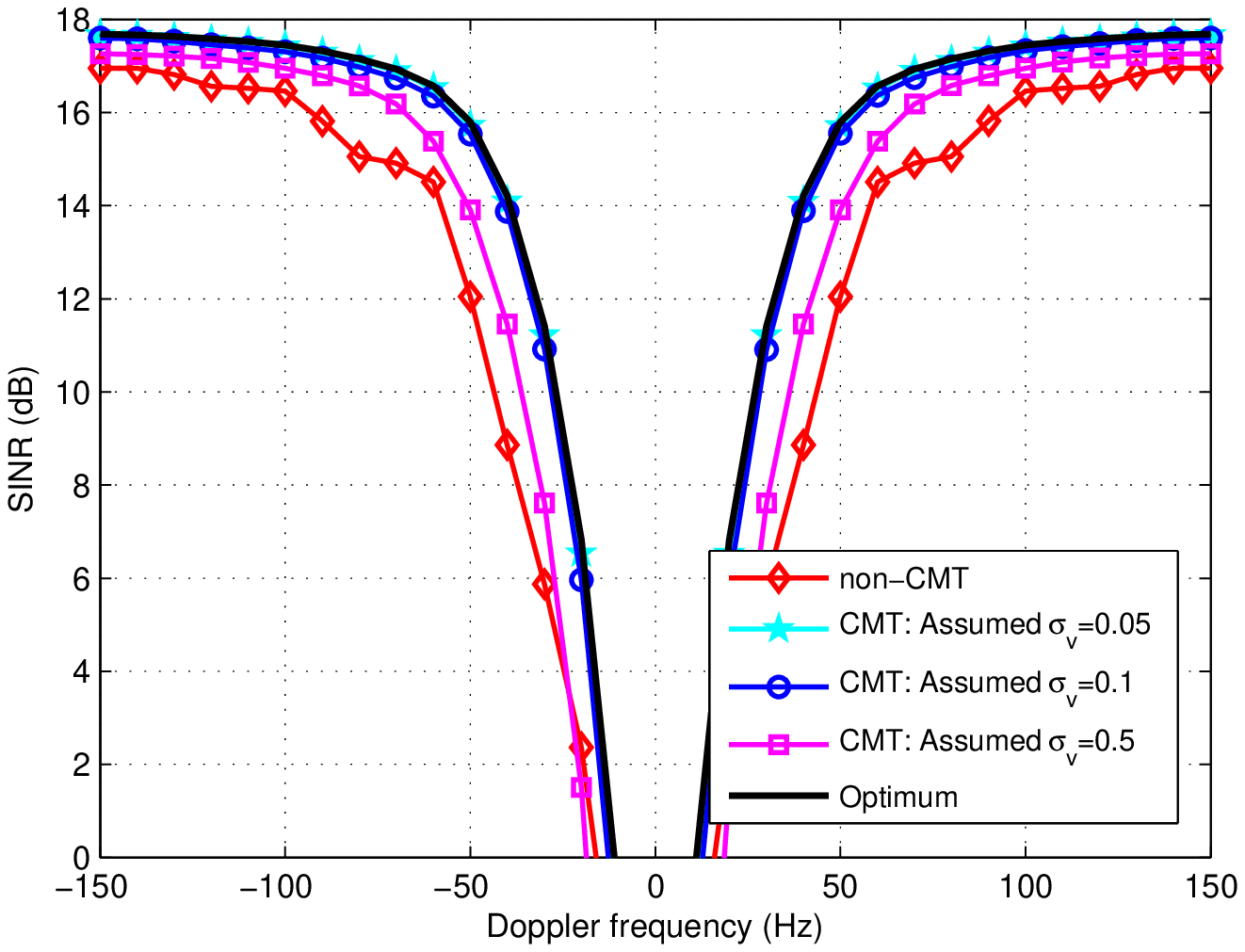}}
  %\hspace{0.5in}
  \subfigure[]{\label{icm.sub3}
  \includegraphics[width=7.8cm]{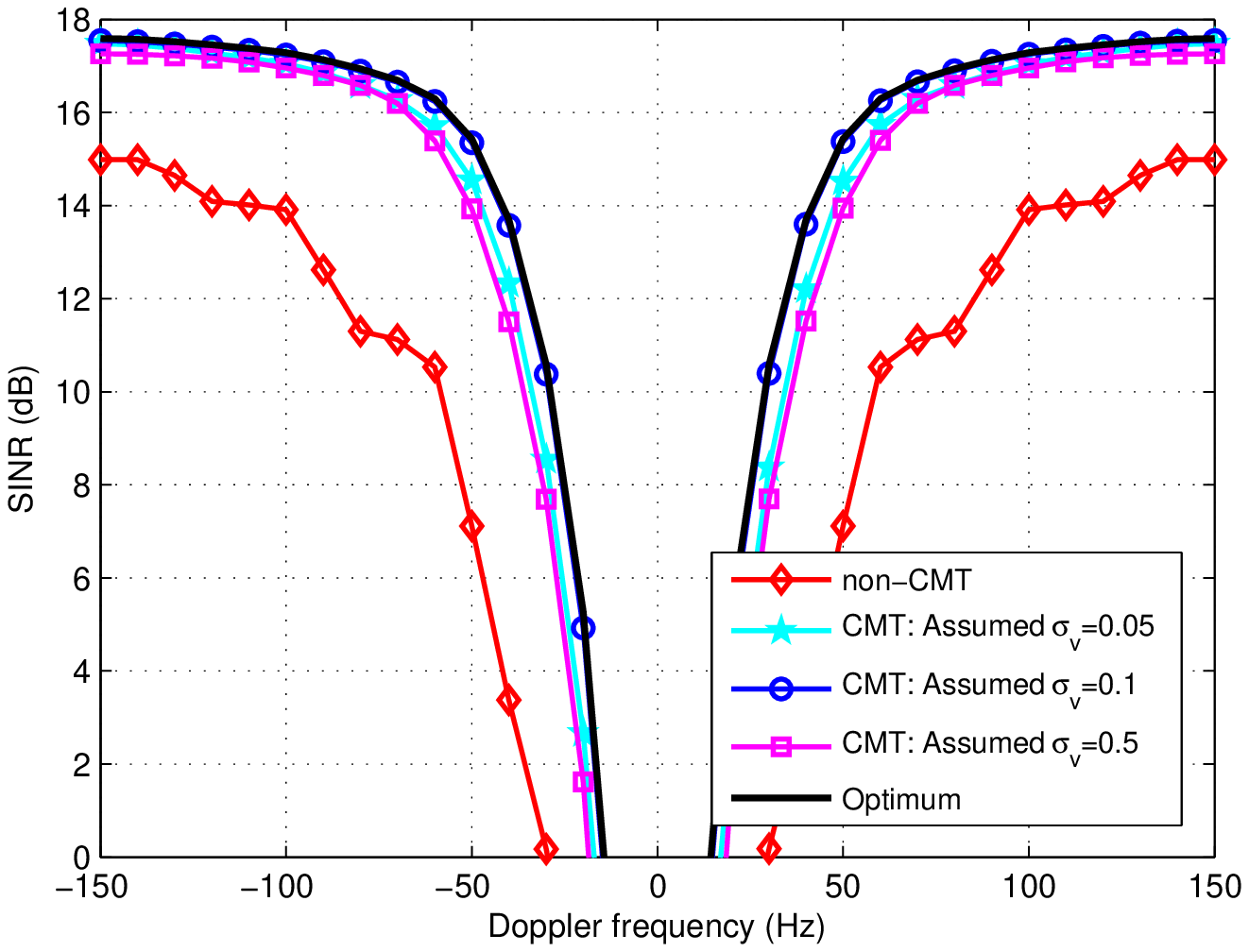}}
  \hspace{0.5in}
  \subfigure[]{\label{icm.sub3}
  \includegraphics[width=7.8cm]{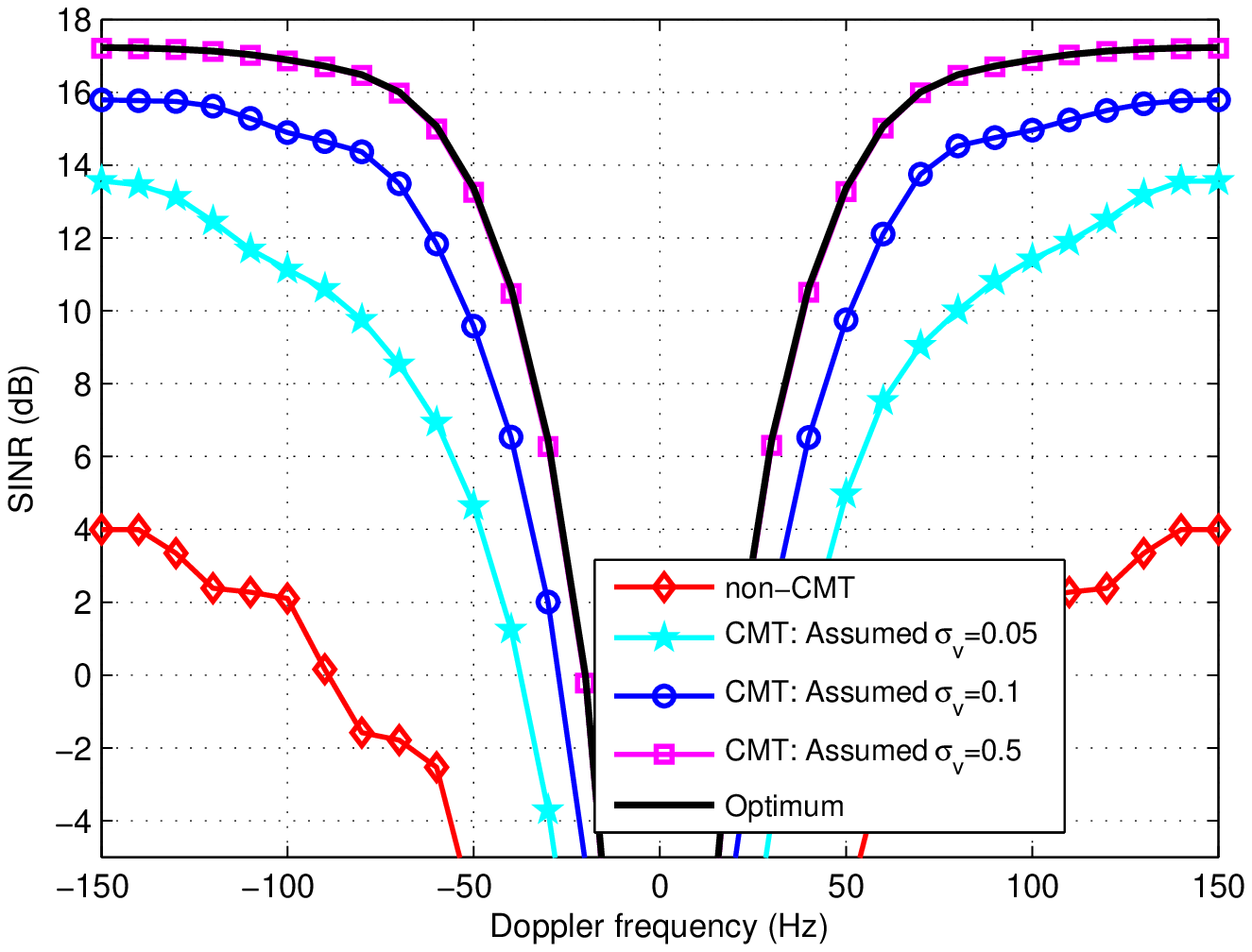}}
  \caption{Impacts of ICM on SINR performance against Doppler frequency with $4$ snapshots and the target Doppler frequency space from $-150$ to $150$Hz. (a): $\sigma_v=0$; (b): $\sigma_v=0.05$; (c): $\sigma_v=0.1$; (d): $\sigma_v=0.5$.}\label{impactsICM}
\end{figure*}

\subsection{Impact of ICM on the SINR Performance}

In this subsection, we evaluate the impact on the SINR performance
with different ICM for our proposed algorithms. In the examples, we
consider four different ICM cases with $\sigma_v=0$,
$\sigma_v=0.05$, $\sigma_v=0.1$ and $\sigma_v=0.5$. The number of
snapshots for training is $4$. In Fig.~\ref{impactsICM}(a), (b), (c)
and (d), we show the SINR performance against the target Doppler
frequency of our proposed LRGP KA-STAP algorithm both with and
without a CMT. From the figures, we observe the following
conclusions. (i) When there is non-ICM, the proposed LRGP KA-STAP
algorithm without a CMT can obtain the optimum performance since the
computed clutter subspace is exact. However, it degrades the SINR
performance with the increase of $\sigma_v$ resulting in extra
sensitivity to the ICM. That is because the computed clutter
subspace can not represent the true clutter subspace. (ii) Our
proposed LRGP KA-STAP algorithm with a CMT illustrates a robust
characteristic to the ICM. When the estimated parameter $\sigma_v$
of CMT is correct, we can achieve the optimum SINR performance.
Furthermore, it is demonstrated the range of values of CMT mismatch
in which the estimated spreading exhibit acceptable SINR
performance, which can be useful in applications. This can be
interpreted as that the computed clutter subspace via the
application of the CMT to the ideal clutter subspace, spans a
similar space to the true clutter subspace.

\subsection{Impact of Inaccurate Prior Knowledge on the SINR Performance}

\begin{figure*}[ht]
  \centering
  \subfigure[]{\label{v.sub1}
  \includegraphics[width=5.5cm]{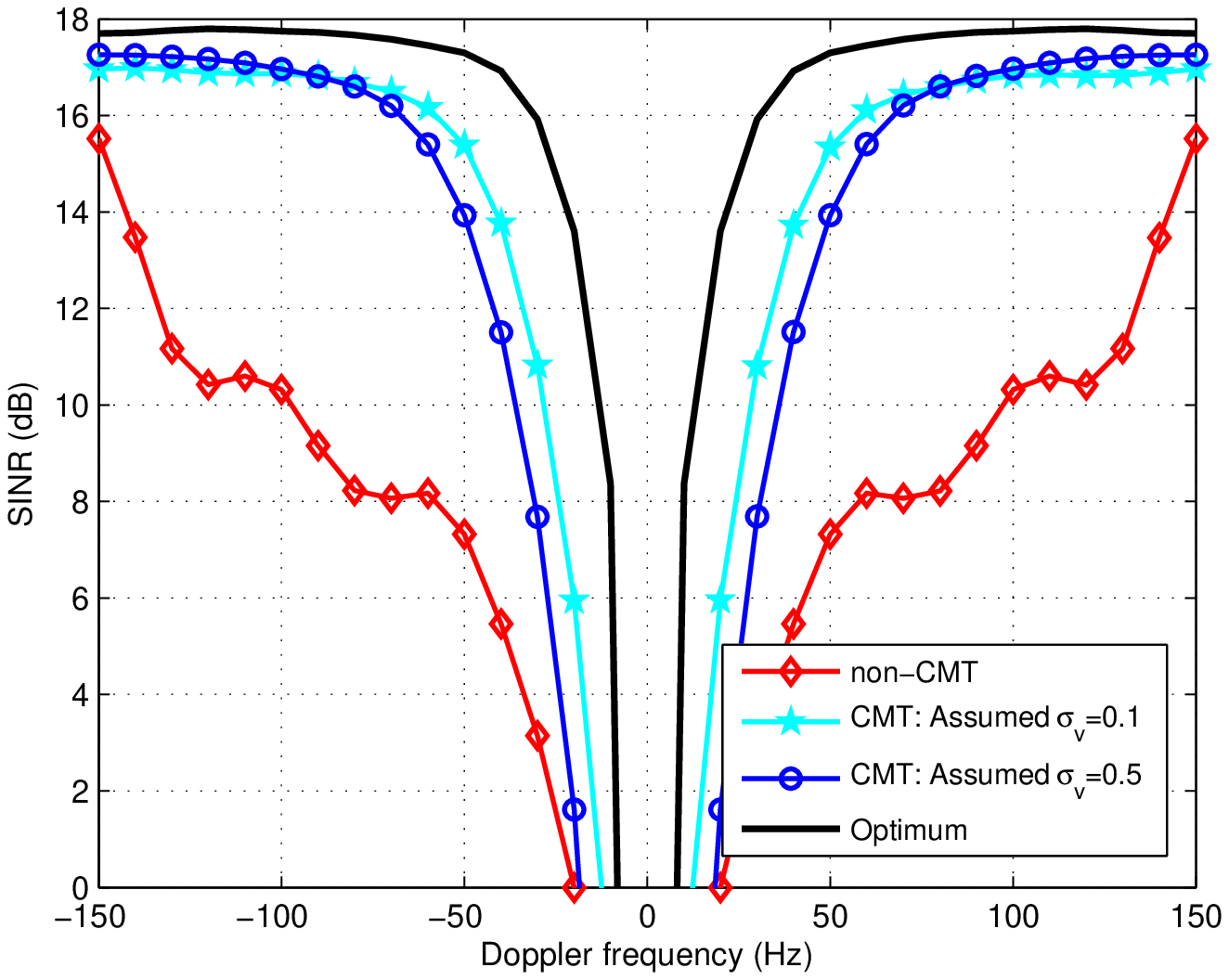}}
  %\hspace{0.4in}
  \subfigure[]{\label{v.sub2}
  \includegraphics[width=5.5cm]{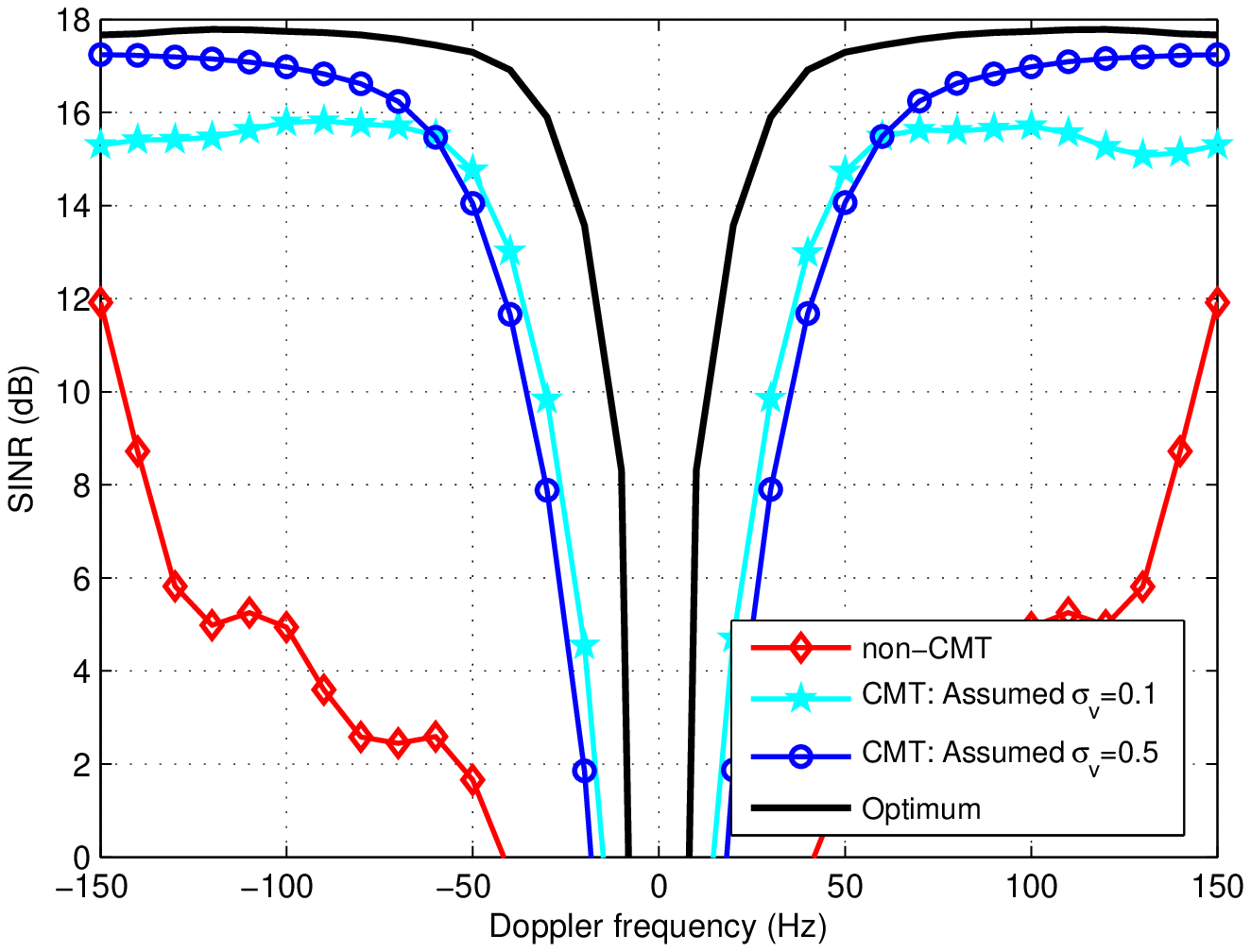}}
  %\hspace{0.4in}
  \subfigure[]{\label{v.sub3}
  \includegraphics[width=5.5cm]{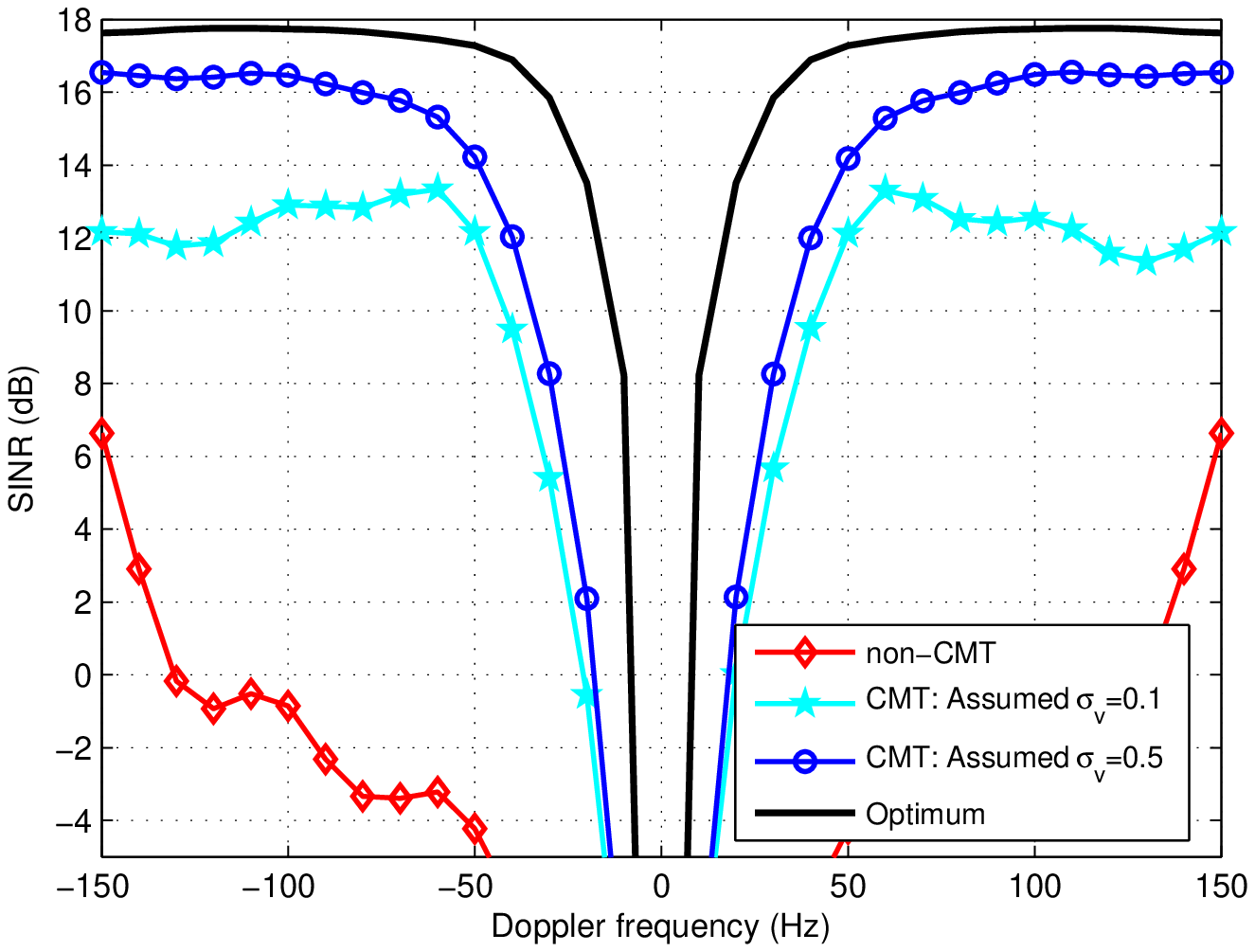}}
  \caption{Impacts of velocity misalignment of the prior knowledge on SINR performance against Doppler frequency with $4$ snapshots and the target Doppler frequency space from $-150$ to $150$Hz. (a): velocity misalignment $0.5$m/s; (b): velocity misalignment $1$m/s; (c): velocity misalignment $2$m/s.}\label{impactsVelocity}
\end{figure*}

In this subsection, we focus on the impact of inaccurate prior
knowledge on the SINR performance of our proposed algorithms. In the
first example, we consider the impact of the velocity misalignment
by showing the SINR performance against the target Doppler
frequency, as shown in Fig.\ref{impactsVelocity}. Consider three
different cases: the velocity misalignments of prior knowledge are
(a) $0.5$m/s; (b) $1$m/s; (c) $2$m/s, compared with true platform
velocity. The potential Doppler frequency space from $-150$ to
$150$Hz is examined and $4$ snapshots are used to train the filter
weights. The plots show that the proposed LRGP KA-STAP algorithm
without a CMT is sensitive to the velocity misalignment, while the
LRGP KA-STAP algorithm with a CMT is robust to that. The reason for
this is that the velocity misalignment of prior knowledge will lead
to the mismatch between the computed clutter subspace and the true
clutter subspace. Although the computed clutter subspace via the CMT
can not avoid this situation, it can mitigate this impact. Because
the velocity misalignment between the clutter patches and the
platform can be seen as the Doppler spreading of the clutter
patches. Moreover, the results also show that a slightly larger
value of the estimated parameter $\sigma_v$ will result in an
improved SINR performance for the velocity misalignment case.

\begin{figure*}[ht]
  \centering
  \subfigure[]{\label{angle.sub1}
  \includegraphics[width=5.5cm]{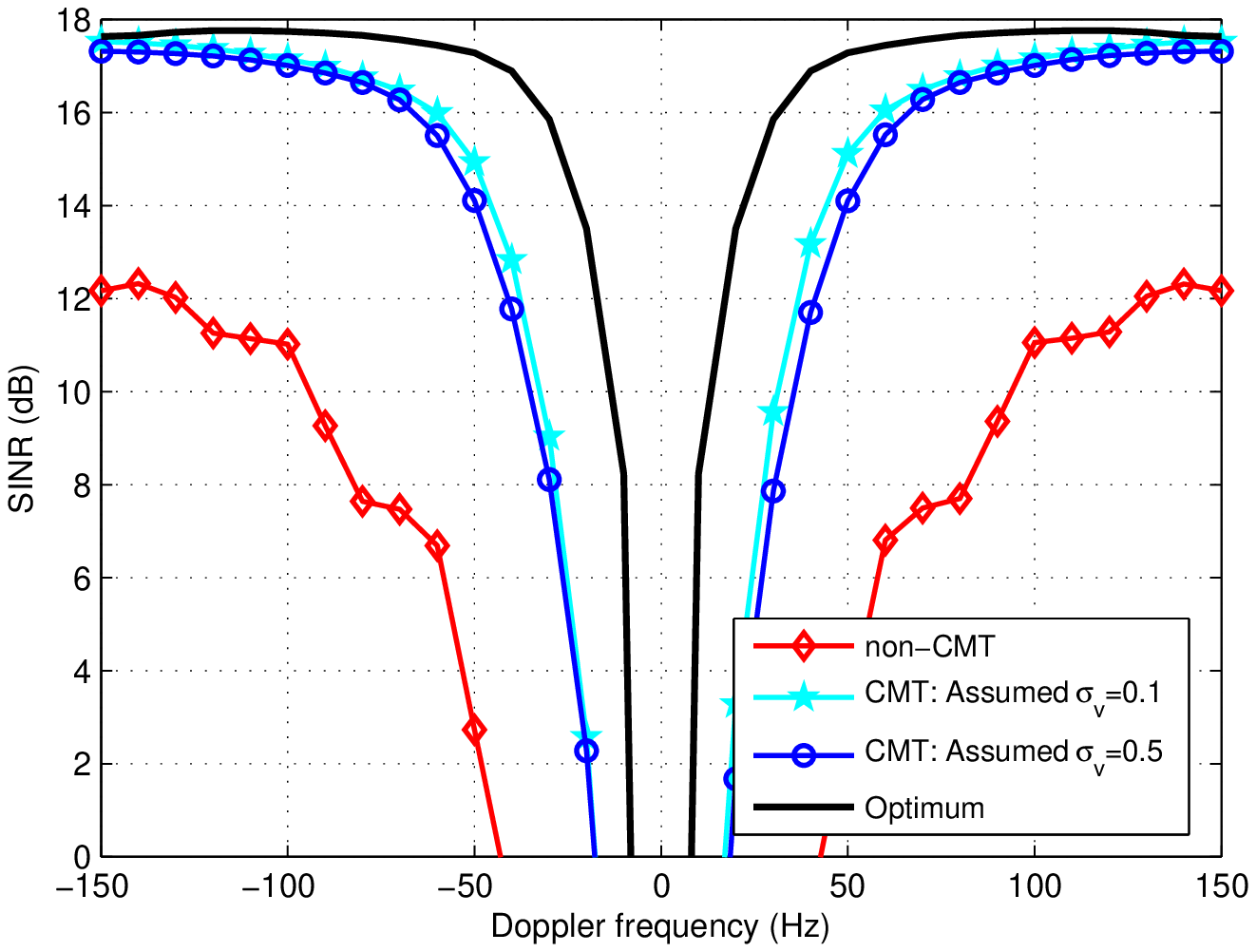}}
  %\hspace{0.4in}
  \subfigure[]{\label{angle.sub2}
  \includegraphics[width=5.5cm]{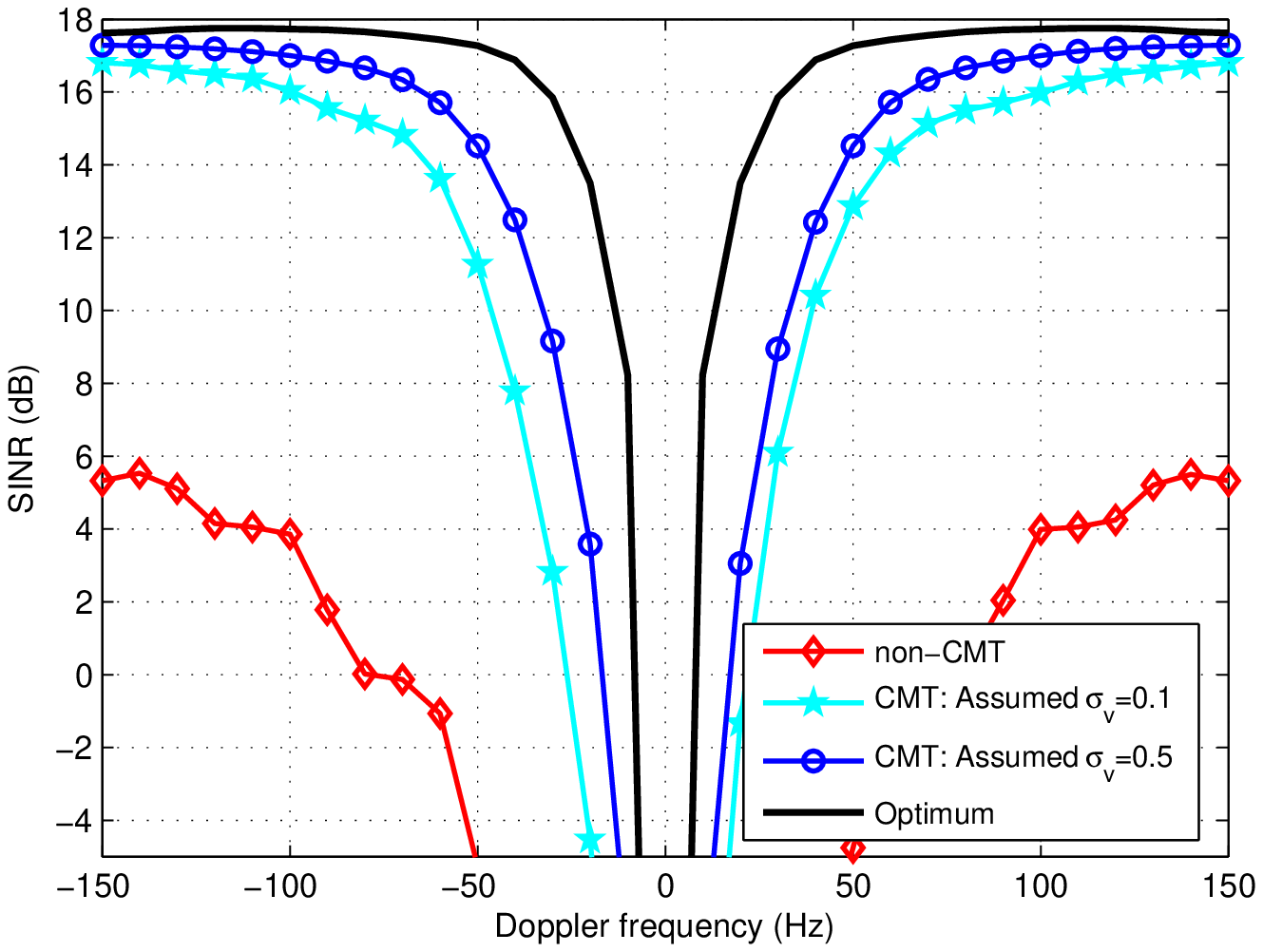}}
  %\hspace{0.4in}
  \subfigure[]{\label{angle.sub3}
  \includegraphics[width=5.5cm]{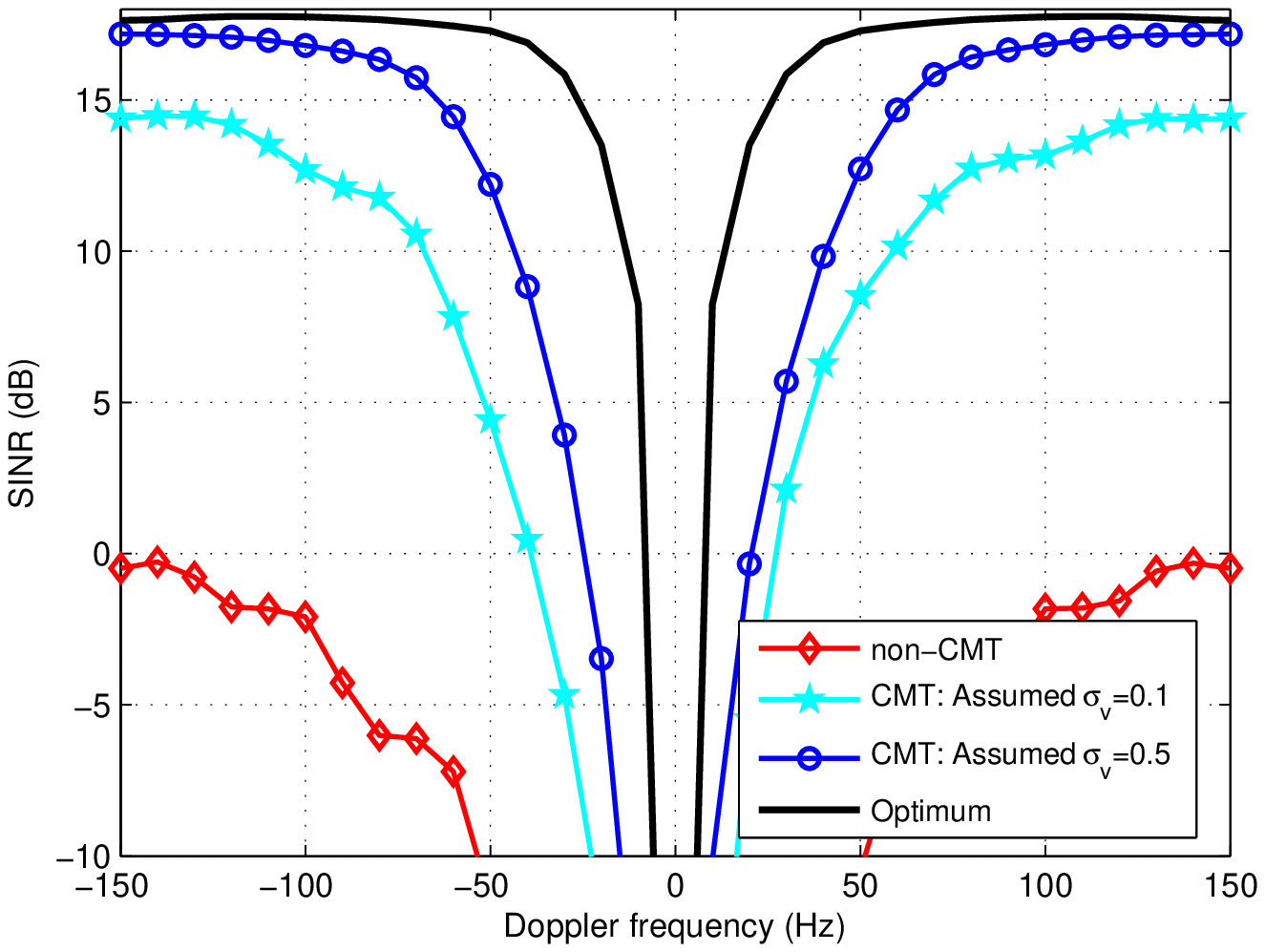}}
  \caption{Impacts of yaw angle misalignment of the prior knowledge on SINR performance against Doppler frequency with $4$ snapshots and the target Doppler frequency space from $-150$ to $150$Hz. (a): yaw angle misalignment $0.2^\circ$; (b): yaw angle misalignment $0.5^\circ$; (c): yaw angle misalignment $1^\circ$.}\label{impactsAngle}
\end{figure*}

The evaluation of the impact caused by the yaw angle misalignment is
shown in Fig.\ref{impactsAngle}, where we also consider three
different cases: the yaw angle misalignments of prior knowledge are
(a) $0.2^\circ$; (b) $0.5^\circ$; (c) $1^\circ$. The curves also
indicate that: (i) the proposed LRGP KA-STAP algorithm without a CMT
is sensitive to the yaw angle misalignment, while the LRGP KA-STAP
algorithm with a CMT is robust to that; (ii) a slightly larger value
of the estimated parameter $\sigma_v$ will result in an improved
SINR performance. The misalignment of the yaw angle will lead to a
Doppler frequency mismatch between the radar platform and the
clutter patches. While the CMT mainly aims at mitigating the
performance degradation caused by the clutter Doppler spreading, the
CMT will lead to an improved estimated clutter subspace and will
exhibit robustness against the yaw angle misalignment.

\subsection{Comparison With Conventional STAP Algorithms}

\begin{figure}[!htb]
\centering
\includegraphics[width=80mm]{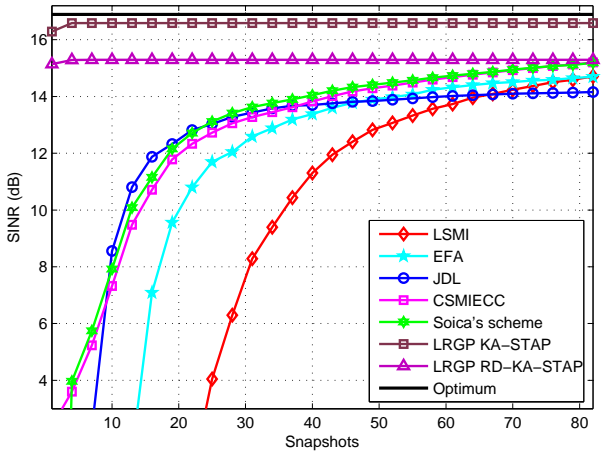}
\caption{
 SINR performance against the number of snapshots considering ICM, where $\sigma_v=0.5$.} \label{sinr_snapshots}
\end{figure}

To provide further investigation about the performance of our
proposed algorithms, we compare the SINR performance versus the
snapshots of our proposed LRGP KA-STAP and LRGP RD-KA-STAP
algorithms with the Loaded SMI (LSMI), the EFA algorithm ($3$
Doppler bins), the $3 \times 3$ JDL algorithm,
 {red}{Stoica's scheme in \cite{Stoica2008} (the prior
knowledge covariance matrix is computed in the same way as the
CSMIECC  {blue}{algorithm}),} and the CSMIECC algorithm (the
combination parameter is set to $0.6$) in \cite{Xie2011}, where the
simulation results are shown in Fig. \ref{sinr_snapshots}. Here, we
consider a scenario of ICM with $\sigma_v=0.5$, and assume the
diagonal loading factors for all algorithms are set to the level of
the thermal noise power. The parameter $\sigma_v$ for our proposed
algorithms is  {red}{assumed to} $1$. The curves in the figure
illustrate that our proposed algorithms have a very fast SINR
convergence speed which only needs three snapshots for training, and
offer significant better SINR steady-state performance compared with
the LSMI, EFA, JDL,  {red}{Stoica's scheme} and CSMIECC algorithms.
This is because the proposed algorithms provide a much better
estimation of the CCM by using prior knowledge of the data, the low
clutter rank property, the geometry of the array and the
interference environment. It should be noted that the SINR
performance of the LRGP RD-KA-STAP algorithm is worse than that of
LRGP KA-STAP with full-DOFs. This is due to the fact that the
reduced DOFs will lead to lower computational complexity at the cost
of performance degradation.

\begin{figure}[!htb]
\centering
\includegraphics[width=80mm]{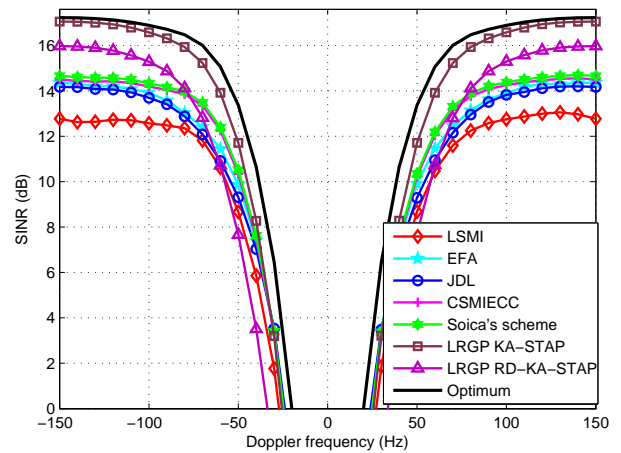}
\caption{ SINR performance versus the target Doppler frequency. The
number of snapshots used for training in the LSMI, EFA, JDL and
CSMIECC algorithms is set to $48$, while we only use $4$ snapshots
for our proposed algorithms.} \label{sinr_doppler}
\end{figure}

The results in Fig.\ref{sinr_doppler} illustrate the SINR
performance versus the target Doppler frequency. The number of
snapshots used for training in the LSMI, EFA, JDL,
 {red}{Stoica's scheme} and CSMIECC
algorithms is set to $48$, while $4$ in our proposed algorithms. It
is found that our proposed LRGP KA-STAP algorithm provides the best
SINR performance among all algorithms, and forms the narrowest
clutter null resulting in improved performance for the detection of
slow targets. It is also shown that the performance of the proposed
LRGP RD-KA-STAP algorithm is worse than that of LRGP KA-STAP with
full-DOFs, but better than other algorithms in most Doppler bins.
Note that although the LRGP RD-KA-STAP algorithm performs slightly
worse than other algorithms in Doppler range of $-60$ to $60$Hz, it
requires much smaller snapshots for training filter weights.

\begin{figure}[!htb]
\centering
\includegraphics[width=80mm]{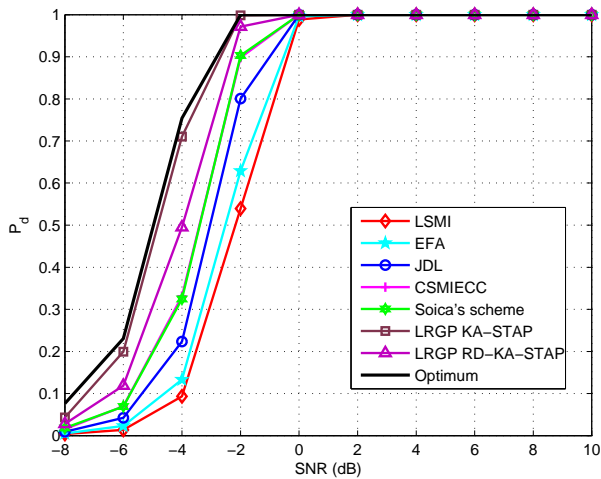}
\caption{ Probability of detection performance against the target
SNR. Suppose the target is injected in the the boresight with
Doppler frequency $100$Hz, and other parameters setting for all
algorithms are the same as that in the second example.}
\label{pd_snr}
\end{figure}

In the next example, as shown in Fig.\ref{pd_snr}, we present the
probability of detection performance versus the target SNR for all
algorithms. The false alarm rate is set to $10^{-3}$ and for
simulation purposes the threshold and probability of detection
estimates are based on $10,000$ samples. We suppose the target is
injected in the the boresight with Doppler frequency $100$Hz. We
note that the proposed algorithms provide suboptimal detection
performance using very short snapshots, but remarkably, obtain much
higher detection rate than other algorithms at an SNR level from
$-8$dB to $0$dB.

\section{Conclusions}

In this paper, novel KA-STAP algorithms have been proposed by using
prior knowledge of LRGP to obtain an accurate estimation of the CCM
with a very small number of snapshots. By exploiting the fact that
the clutter subspace is only determined by the space-time steering
vectors, we  {red}{have developed a Gram-Schmidt orthogonalization
approach to compute the clutter subspace. In particular, for a
side-looking ULA,} we have proposed a scheme to directly select a
group of linearly independent space-time steering vectors to compute
the orthogonal bases of the clutter subspace. Compared with the LSE
algorithm, it has not only exhibited a low complexity, but also
shown a simple way to compute the CCM. To overcome the performance
degradation caused by the non-ideal effects
 {red}{and the prior knowledge uncertainty}, the proposed
KA-STAP algorithm that combines the CMT has been presented and a
reduced-dimension version has been devised for practical
applications.  {blue}{This has also provided evidence that is
feasible to directly use the received data vector and the calibrated
space-time steering vectors (only the spatial taper without the
temporal taper) to compute the assumed clutter amplitude.} The
simulation results have shown that our proposed algorithms
outperform other existing algorithms in terms of SINR steady-state
performance, SINR convergence speed and detection performance for a
very small number of snapshots, and also exhibit robustness against
errors in prior knowledge.

%\bibliographystyle{IEEEbib}
%\bibliography{refs_jst}

\end{document}